\documentclass{article}

\usepackage[preprint]{neurips_2025}

\usepackage[utf8]{inputenc} 
\usepackage[T1]{fontenc}    
\usepackage{CJKutf8}
\usepackage{hyperref}       
\usepackage{url}            
\usepackage{xurl}
\usepackage{tabularx}
\usepackage{booktabs}       
\usepackage{amsfonts}       

\usepackage{svg}
\usepackage{nicefrac}       
\usepackage{microtype}      
\usepackage[table,xcdraw]{xcolor}         
\hypersetup{hidelinks} 

\usepackage{subcaption}
\usepackage{wrapfig}
\usepackage{multirow}
\usepackage{makecell}
\usepackage{xspace}
\usepackage{enumitem}
\usepackage{amsmath}
\usepackage{listings}
\usepackage{algorithm}
\usepackage{algorithmic}
\usepackage{amsfonts,amssymb}
\usepackage{amsthm}
\usepackage{mathrsfs}
\usepackage{subcaption}
\usepackage{natbib}
\setcitestyle{numbers,square}
\usepackage{graphicx}       
\usepackage{cleveref}
\usepackage{fontawesome}
\usepackage{utfsym}
\usepackage[most]{tcolorbox}
\usepackage{siunitx} 
\usepackage{tabularx}
\usepackage{CJKutf8}
\usepackage{fancyvrb}
\usepackage{colortbl}
\usepackage{array}
\usepackage{threeparttable}
\usepackage{CJKutf8}
\usepackage{CJK}
\usepackage{booktabs}
\usepackage[table]{xcolor}
\usepackage{lipsum}
\usepackage{listings}
\usepackage{multicol}

\definecolor{demphcolor}{RGB}{144,144,144}

\definecolor{linkblue}{RGB}{0, 112, 192}

\usepackage{hyperref}
\hypersetup{
    colorlinks=true,
    linkcolor=linkblue,  
    citecolor=linkblue,  
    urlcolor=linkblue,   
    filecolor=linkblue
}

\newcommand{\hhide}[1]{}
\newcommand{\hide}[1]{}

\definecolor{zhipublue}{HTML}{3859FF}

\newtcolorbox{promptbox}[1][]{
  breakable,
  title=#1,
  colback=gray!5,
  colframe=black,
  colbacktitle=gray!15,
  coltitle=black,
  fonttitle=\bfseries,
  bottomrule=1.5pt,
  toprule=1.5pt,
  leftrule=1pt,
  rightrule=1pt,
  arc=0pt,
  outer arc=0pt,
  enhanced,
  before upper={\parindent=1.5em} 
}

\theoremstyle{definition}

\title{   \Large StarCore: The Model that Dreams the Cybersecurity World}

 \author{\raisebox{-0.3em}{\includegraphics[height=1.2em]{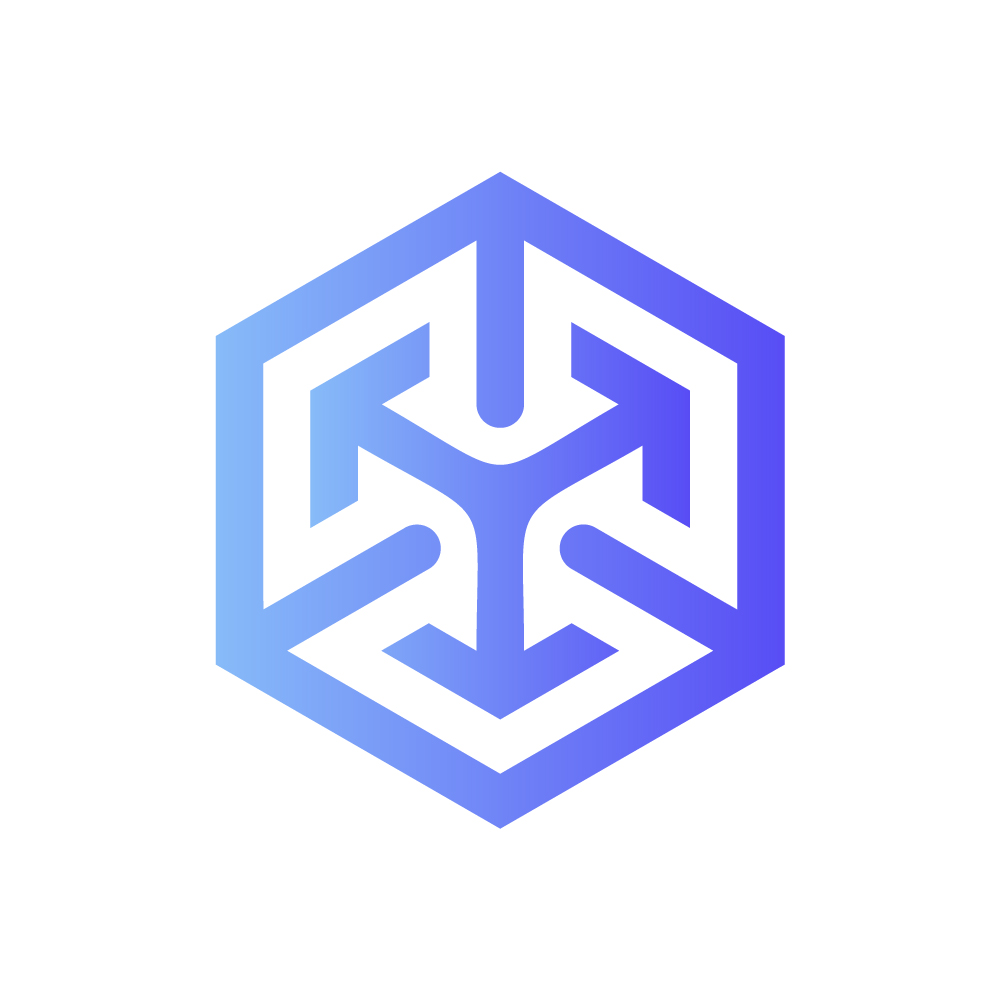}}
{StarCore Team}
~\\\\
Metaordo ~\&~ Tsinghua University\\\\
(For the complete list of authors, please refer to the \hyperref[sec:contribution]{Contribution} section)
{}
}

\begin{document}

\maketitle

\vspace{-1.5em}
\begin{abstract}

\vspace{-0.5em}
We present StarCore, a next-generation world model that extends the world-model paradigm to cybersecurity. Building on the ability to learn multimedia data in system environments, including code, commands, traffic, and agent traces, StarCore adopts system state chain learning to significantly improve performance in understanding and analyzing dynamic data streams across diverse network systems. To advance models for dynamic systems, we implement a new state-chain learning framework for system data that models system states with diverse benign and malicious behaviors and introduces three state-chain supervised fine-tuning tasks for system-world understanding. Furthermore, we propose a novel RL framework based on GRPO algorithms with state-chain rewards that further improves understanding of system state chains and reasoning for security analysis. With these innovations, StarCore achieves state-of-the-art performance on state-chain completion, prediction, and security analysis tasks. Most critically, StarCore demonstrates unprecedented capability in understanding dynamic real-world system information streams, surpassing traditional foundation models that rely on simple QA reasoning for cybersecurity. Code, models, and more information are available at \url{https://github.com/metaordo/StarCore}.

\end{abstract}

\begin{figure}[!h]
    \centering
    \includegraphics[width=\linewidth]{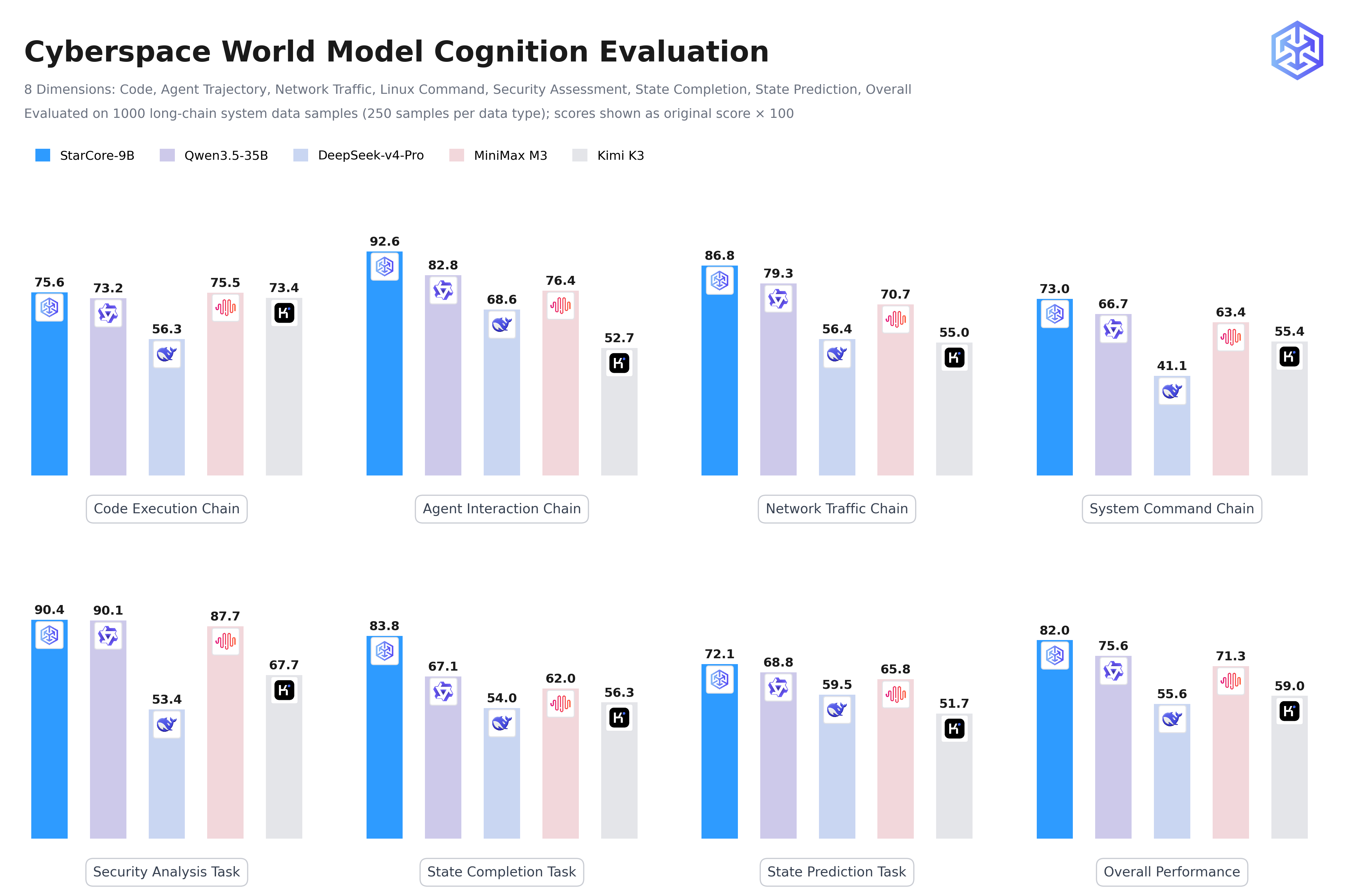}
    \caption{Results of StarCore-9B, Qwen3.5-35B, DeepSeek-v4-Pro, Minimax M3, and Kimi K3 on 8 cyberspace world cognition benchmarks: Code Execution Chain, Agent Interaction Chain, Network Traffic Chain, System Command Chain, Security Analysis Task, State Completion Task, State Prediction Task, and Overall Performance Benchmark.}
    \label{fig:arc}
\end{figure}

\section{Introduction}
With the rapid development of the Internet and artificial intelligence, cyberspace has evolved from a traditional information-exchange environment into a complex, dynamic system world composed of software code, operating systems, network protocols, business applications, and autonomous agents. In parallel, cyber attacks have become increasingly stealthy, intelligent, and sophisticated. Attack behaviors are no longer confined to a single vulnerability exploit or a fixed attack pattern. Instead, they have evolved into multi-stage campaigns that unfold continuously around the target system's running state.

Mainstream cyber-defense technologies rely on human-written rules, signature matching, and machine-learning classifiers that analyze a single data type at a single point in time. For instance, Intrusion Detection Systems (IDS)~\cite{FlowLens}\cite{HyperVision}\cite{Poseidon} recognize anomalous patterns in network traffic with predefined rule bases or learned models. Security Information and Event Management (SIEM)~\cite{market} platforms correlate logs, alerts, and events across hosts. Provenance-based threat-hunting systems reconstruct attack campaigns by correlating suspicious information flows and kernel audit records~\cite{milajerdi2019holmes}. Although effective within their own scopes, these approaches treat each data modality as an isolated island, producing static entity-relationship snapshots rather than a model of how system states transition over time.

To enhance the intelligence and automation of network security analysis, large language models~\cite{mistral}~\cite{llama3}~\cite{deepseek}~\cite{qwen3}~\cite{language-model} are introduced to reshape security operations. Microsoft Security Copilot~\cite{microsoft} combines an LLM with security data sources and a tool-calling orchestration layer to analyze alerts and explain threats. Cisco~\cite{Cisco} continually pre-trains a Llama-3.1~\cite{llama3} backbone on CVE, CWE, ATT\&CK, and threat-intelligence corpora to obtain a security-knowledge reasoning model. Despite strong knowledge comprehension and textual reasoning capabilities, these models are trained on static text, and their knowledge representation is dominated by semantic relations among documents. They can answer what a vulnerability affects or summarize what an attack log contains, but cannot perceive how an attack action alters subsequent system states, or how one running state transits into another. In short, they know security without understanding the system world.

\paragraph{Design.} We present StarCore, a cybersecurity world model built upon system state chain learning. StarCore treats the system data streams within a network system world as the minimal set of data needed for understanding. The heterogeneous modality data produced during system operation, including execution code, system commands, network traffic, and agent interaction trajectories, are treated as world-state descriptions. Within each modality, consecutive states are chained in system execution order, forming large-scale, long-distance directed execution state chains that structurally represent the system world and capture attack propagation through system-state evolution. Through a new state-chain learning mechanism, StarCore models the system world’s states with diverse benign and malicious behaviors and introduces novel state-chain learning tasks using SFT and RL for system-world understanding. It demonstrates a concrete path from knowledge-driven security question answering toward world-model-driven security understanding and reasoning, enabling attack-chain tracing, attack-evolution prediction, and dynamic risk analysis for proactive defense.

\paragraph{Methods.} StarCore instantiates a capability progression from seeing the system world to understanding it. First, the model undergoes continued post-training on security knowledge to internalize vulnerability, attack, and threat-intelligence knowledge. Second, we perform state-chain supervised fine-tuning (SFT) to learn state-transition regularities through three state-chain learning tasks, including state chain completion, state chain prediction, and state chain security analysis, combined with explicit reasoning traces for system state understanding. Third, we align the model with a state-chain reinforcement learning method. Instead of optimizing textual answer matching as in conventional RLHF, we take the correctness of state-chain reasoning as the reward basis, design a multi-dimensional State Chain Reward (SCR) that scores output format, thinking quality, and answer correctness to optimize the policy with Group Relative Policy Optimization (GRPO)~\cite{shao2024deepseekmath}. In summary, the leap from security knowledge to world understanding in StarCore is driven by the following technical contributions:

\begin{itemize}[leftmargin=1.5em,itemsep=2pt]
\item First, we propose a state-chain modeling method for the network system world. StarCore defines a unified state space over execution code, traffic flows, system commands, and agent traces. These system execution state chains are constructed to transform static and isolated event records into coherent and temporally ordered behavioral stories of the system world.

\item Second, we present a state-chain-supervised world-understanding learning paradigm. Rather than learning question-answer mappings over security knowledge, StarCore is supervised directly on system state chains through completion, prediction, and security-analysis tasks, with large-scale long-distance state reasoning for world understanding, thereby internalizing the evolutionary regularities, causal mechanisms, and cross-modality associations of network system states.

\item Third, we introduce a state-chain-based reinforcement learning approach to align and enhance cybersecurity reasoning capability. StarCore uses a multi-dimensional State Chain Reward (SCR) method to evaluate system state-chain understanding across modalities and employs the GRPO algorithm to explore diverse reasoning paths and possible states in running systems, substantially improving generalization to unseen attack variants and complex dynamic environments.
\end{itemize}

With these advancements, StarCore challenges traditional foundation model paradigms that rely on simple QA reasoning for cybersecurity and serves as a world model that captures dynamic, real-world system information streams for security-task reasoning. We release StarCore to the community to further advance the frontier of cybersecurity foundation models.

\section{System State-Chain Modeling}
The first pillar of StarCore is a state-chain modeling method that turns raw, heterogeneous system data into state chains of the system world. We proceed in two steps: constructing a unified system state space and organizing states into system state chains.

\subsection{A Unified State Space for the System World}

\paragraph{System World.} The system world is the minimal set of system data streams, such as execution codes, network traffic, and system commands, together with the interaction trajectories of agents running on top of them. It comprises the entities, behaviors, and environmental changes involved in operating a network system. 

\paragraph{World Model.} The world model~\cite{ha2018worldmodels}~\cite{lecun2022autonomous}~\cite{hafner2025dreamerv3}~\cite{zuo2026qwenagentworld}~\cite{chen2026codeworldmodel}~\cite{fair2025cwm} aims to develop a foundational model capable of comprehending the operational states and interrelationships within the realm of network systems, while predicting future attack behaviors. The model will construct and interpret the dynamic environment formed by aggregating system data and leverage this understanding to analyze and forecast future attacks.

\paragraph{System State Chain.} To achieve such capability, we define the data-chain scope of the system world to cover four key modalities that span the complete behavioral spectrum from low-level binary execution to high-level agent interaction, including the chain of code execution, network communications, system command execution, and agent behaviors. To model the state-transition process of the dynamic network system world, we first formulate the run-time data of the four modalities as system states and then represent the complete state-transition process as a state chain, thereby capturing massive normal and malicious system behaviors throughout the system world.

\begin{figure}[!t]
    \centering
    \includegraphics[width=\linewidth]{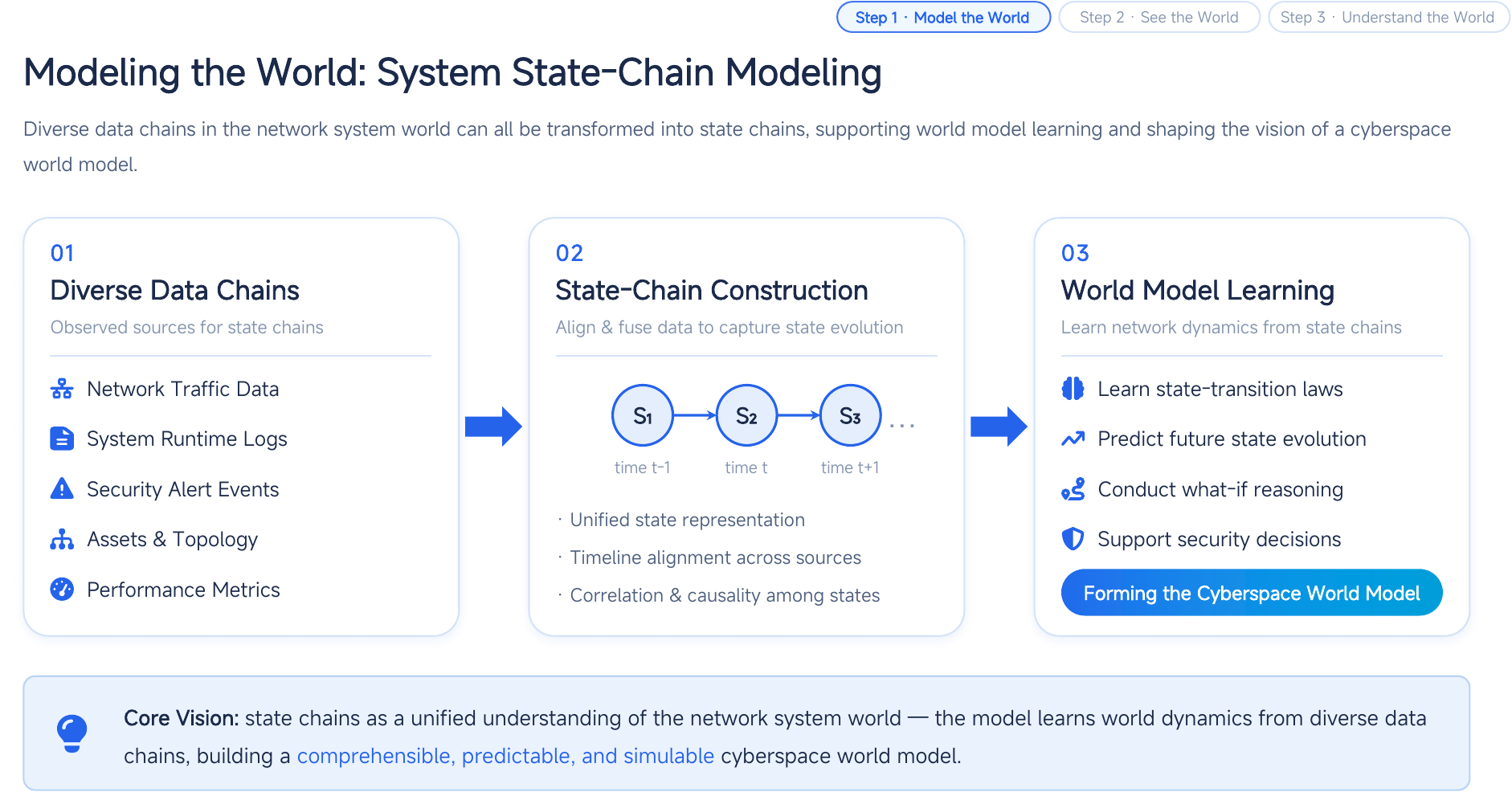}
    \caption{\textbf{Modeling the world:} StarCore models the data transition knowledge observed in the running systems as the system state chains to build the world view of the cyberspace.}
    \label{fig:data-modeling}
\end{figure}

Given the system data that represents the current state of the system ${o_t}$ at the time step $t$, which comes from the last state of system data $o_{t-1}$ with the benign or malicious system action $a_{t-1}$, where $a_t \in \mathcal{A}$ denotes the system action space that running behind the observed system data, the world model $G_{WM}(o_t, a_t)$ should model the complete system transition trajectory $\tau =\{o_0, a_0, o_1, a_1, ..., o_T\}$. For a time step $t$, we define the system data chain as $c =\{o_0, o_1, ..., o_T\}$ to help the model capture the system running actions $\alpha = \{a_0, a_1, ..., a_{T-1}\}$ with different benign or attack intentions.

\subsection{Uni-modal System State Chain} 

Within a single modality, behaviors are continuous, temporally ordered, and governed by evolutionary regularities. We therefore concatenate consecutive states of the same modality along the time axis to form uni-modal system state chains, each being a complete behavioral sketch of an entity in that modality. In this paper, we use four types of chains shown as follows:
\begin{itemize}[leftmargin=1.5em,itemsep=2pt]
\item \textbf{Code Execution Chain.} Tracing the instruction-execution trajectory of a program or script run, the code states of sequentially executed basic blocks or code lines are concatenated along the control flow, revealing the dynamic path from entry to exit.
\item \textbf{Network Traffic Chain.} Around one network session or a group of related flows, packet or flow states are concatenated in chronological order, characterizing the full communication sequence from handshake and negotiation through data transfer to teardown.
\item \textbf{System Command Chain.} Over the lifecycle of a user session or service process, the log states of executed commands or audit events are concatenated by timestamp, reflecting the operator’s action sequence and evolving intent.
\item \textbf{Agent Interaction Chain.} Within one agent task session, the states of dialog turns or tool-call steps are concatenated, describing the agent’s continuous decisions, observations, and environment interactions.
\end{itemize}

Figure~\ref{fig:data-modeling} shows the details of building the different types of system state chains. We use the system state chain to model the dynamic state transitions that occur in a real-world network system. Through this mapping, originally isolated code fragments, packets, log entries, and dialog turns become synchronized state snapshots in a common space, giving the upper-layer model, for the first time, a cross-modal, panoramic view of the running system.

\subsection{Multi-modal System State Chain}

Once attacks execute across multi-modal chains, complex system interactions usually cannot be captured with statistical QA and the limited windows of traditional LLMs. Therefore, the world model also aims to cover a wide range of system states through multi-modal system interaction chains. For example, a complex attack usually triggers multiple system-state chains simultaneously; remote shell attacks may trigger both system-command chains and network-traffic chains, producing stealthy command traffic when creating a remote shell and exfiltrating data.

\begin{figure}[!t]
    \centering
    \includegraphics[width=\linewidth]{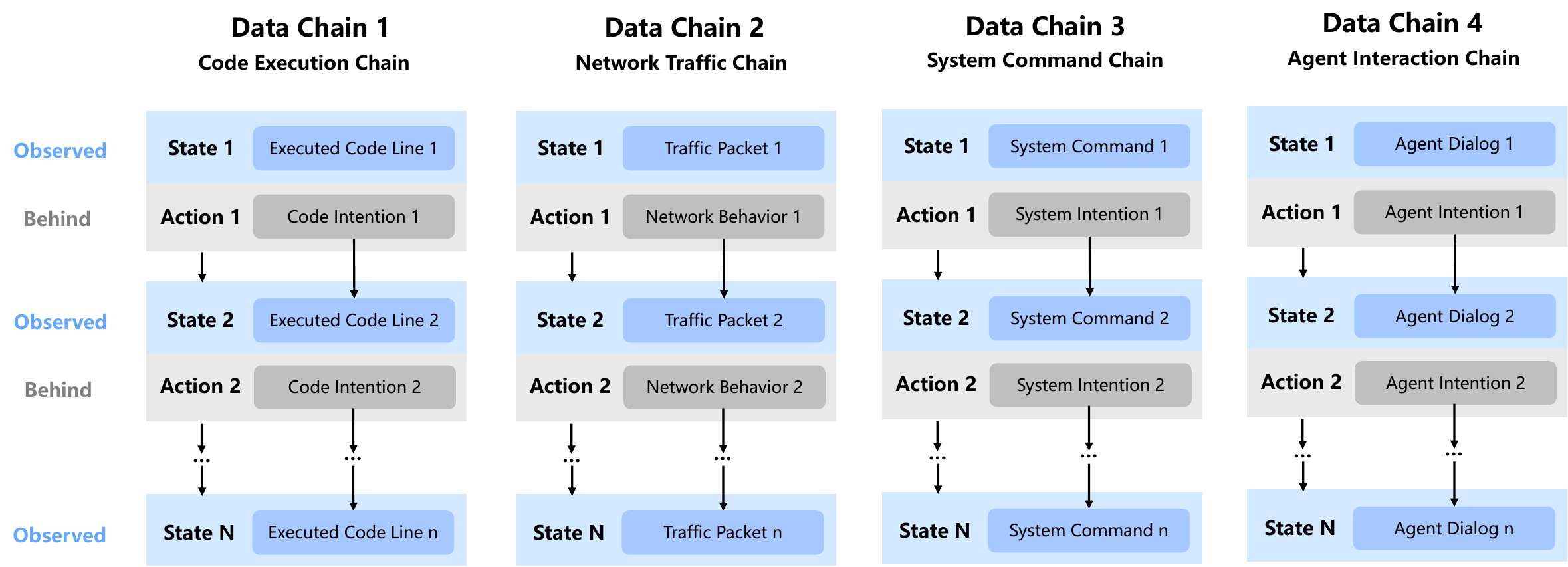}
    \caption{Illustration of the system state-chain construction among four types of data chains.}
    \label{fig:data-chain}
\end{figure}

\textbf{World Prototype: Interaction Graph with System State Chains.} Modeling with large-scale multi-modal system state chains, the world model will face an extra network space of system state chains to form the interaction graph in the cyberspace. Figure~\ref{fig:data-chain} illustrates the interaction graph. When a malicious attack or benign behavior involves multiple system data chains, the equivalent actions across chains produce intersection points as common states that connect two or more system state chains to build the interaction graph of the cyberspace world.

Through this mapping, originally isolated code fragments, packets, log entries, and dialog turns become synchronized state snapshots in a common space, giving the upper-layer model, for the first time, a cross-modal, panoramic view of the running system.

\section{State-Chain Supervision Fine-Tuning}

\begin{figure}[!t]
    \centering
    \includegraphics[width=\linewidth]{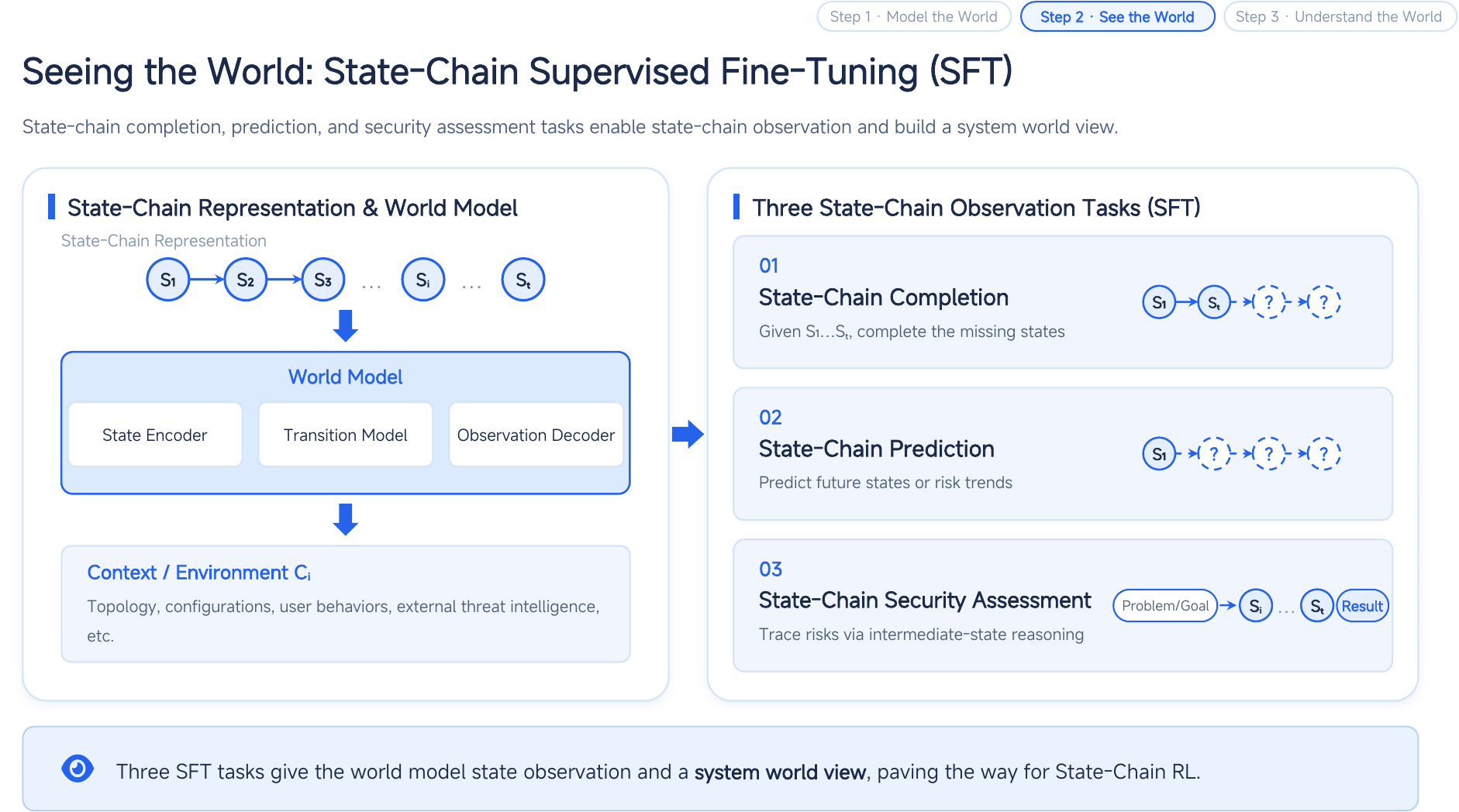}
    \caption{\textbf{Seeing the world:} StarCore receives system state chains by observing the context of the chains through three specific state-chain supervised fine-tuning tasks.}
    \label{fig:sft}
\end{figure}

The supervision fine-tuning phase of StarCore aims to transform traditional LLMs into a highly capable world model with cyberspace data-chain reasoning abilities. As shown in Figure~\ref{fig:sft}, we propose a state-chain supervision fine-tuning (SFT) pipeline that uses system state chains rather than knowledge question-answer pairs as the supervision target, lifting the model from perceiving system states to understanding them. By leveraging this unique training paradigm and three state-chain learning tasks, StarCore achieves system data-chain completion, future event prediction, and detailed security analysis, bridging the reasoning gap in long-chain reasoning in realistic cyberspace. 

\begin{figure}[!t]
    \centering
    \includegraphics[width=\linewidth]{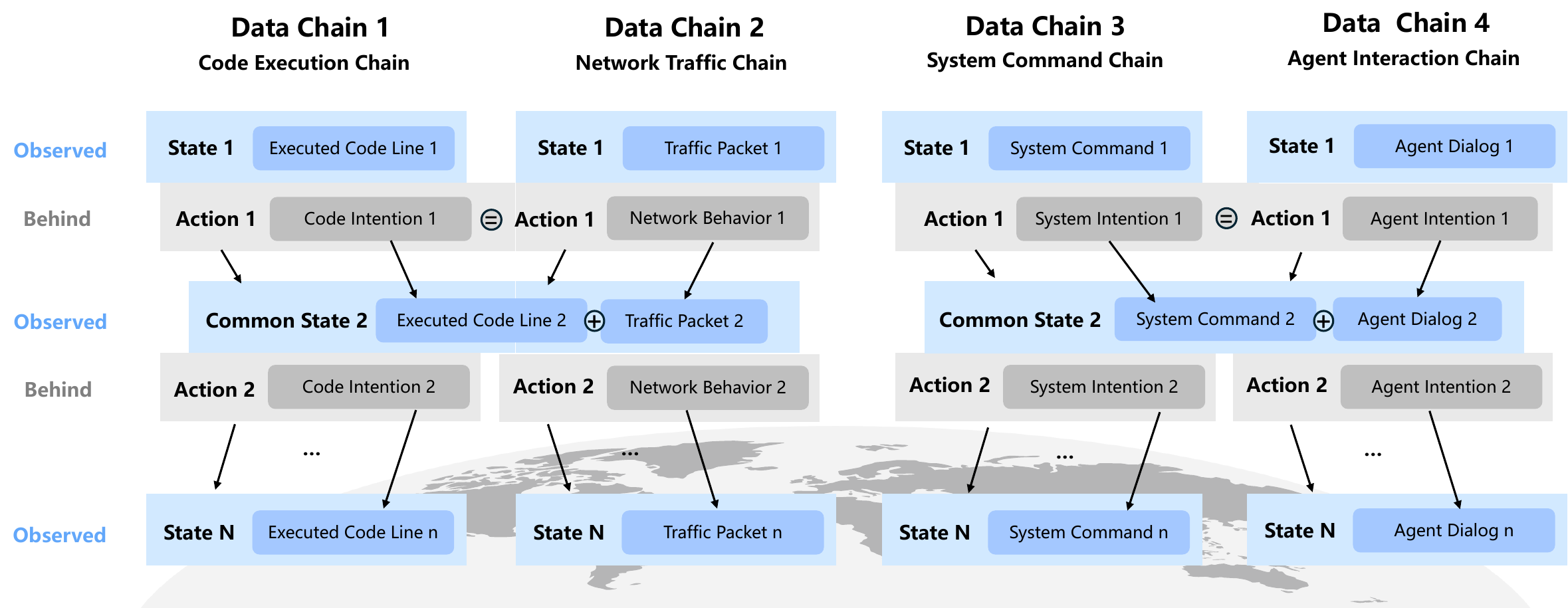}
    \caption{Illustration of the interaction graph built by multi-modal system state chains.}
    \label{fig:data-chain-graph}
\end{figure}

\subsection{Training Paradigm}

Compared with traditional LLMs, StarCore significantly expands the scale of observation and reasoning for the system state chains during the SFT stage. The state-chain SFT data structure of StarCore covers five major categories:
\begin{itemize}[leftmargin=1.5em,itemsep=2pt]
    \item \textbf{Instruction}: The predefined system instruction that describes the task to conduct or the role to play in the cybersecurity area.
    \item \textbf{Question}: The user input that includes the question based on the cyberspace system data.
    \item \textbf{States}: The states are the core system state chains the world model should observe, including code, traffic data, system logs, agent trajectory, and other data that can be modeled as system state chains.
    \item \textbf{Action Reasoning}: The reasoning for the actions is the evidence the world model finds from observing the states, describing the possible actions behind the state sequence, and explaining why the state transition occurred and what the action intents are.
    \item \textbf{Answer}: The answer provides the world model's final decision to solve the complete analysis task queried by users.
\end{itemize}

The SFT corpus of StarCore covers three major categories: open-sourced datasets, self-collected datasets, and AI-generated datasets. StarCore utilizes large-scale security-related QA and CVE/CWE databases to build the foundation knowledge of the world model. To model various state chains in a running system, we collect billions of lines of code, traffic data, command logs, and agent trajectories from publicly available repositories on the Internet, and generate real-world system-running data from controlled cloud servers. To enable state-of-the-art reasoning ability, StarCore uses Qwen3.5-9B as the base model, and learns from reasoning results from well-known LLMs including GLM-5.2~\cite{glm-5}, MiniMax M3~\cite{minimax2026m3}, DeepSeek-V4-Pro~\cite{deepseek2026v4}, and Kimi K3~\cite{kimi-k3} as the reference of action reasoning, which gives the chance to combine frontier LLM knowledge for cybersecurity reasoning.

\subsection{State-Chain Learning Tasks}

During the state-chain learning stage, we teach the world model to observe the system through state chains.  We then construct three supervised fine-tuning tasks through state-chain completion, state-chain prediction, and state-chain security assessment, giving the world model genuine state-observation ability and establishing its system world view.

StarCore employs three state-chain learning tasks to realize state-chain supervised fine-tuning:
\begin{itemize}[leftmargin=1.5em,itemsep=2pt]
    \item \textbf{State Chain Completion}: The state chain completion task requires the world model to determine missing states from context, building an understanding of the full transition.
    \item \textbf{State Chain Prediction}: The state chain prediction task requires the world model to predict the next state based on historical chains, enabling it to anticipate future actions early on.
    \item \textbf{State Chain Security Assessment}: The state chain security assessment task requires the world model to analyze the possible abnormal behaviors or malicious intentions behind the data chains.
\end{itemize}

\subsection{State Chain Completion}
The state chain completion task aims to enable the world model to recover missing system states by understanding the contextual dependencies and evolution patterns within an observed state chain. In real-world network systems, system observations are often incomplete due to packet loss, insufficient logging, partial data collection, or limited visibility into system execution. These missing observations may obscure critical behavioral transitions and break the continuity of the system state chain. We use $S = \{s_0, ..., s_T\}$ to model the system state chains. $C$ represents the context and environment of the system. Therefore, the world model should learn to infer unobserved states from their preceding and succeeding states, rather than simply memorizing individual observations. 

\begin{equation}
\arg\max_{S_{\mathcal{M}}}
P_{\theta}(S_{\mathcal{M}}\mid
\widetilde{S},C),
\end{equation}

where $S_{\mathcal{M}}={s_t\mid t\in\mathcal{M}}$ denotes the ground-truth missing states and $\theta$ represents the parameterized world model. We use $R$ to model the world model's reasoning. $\mathcal{D}$ is the SFT dataset for the state chain completion task.

\begin{equation}
-\mathbb{E}{(S,C,R)\sim\mathcal{D}}
\left[
\log P_{\theta}
(S_{\mathcal{M}},R_{\mathcal{M}}
\mid \widetilde{S},C)
\right].
\end{equation}

Through this task, the world model learns to recover missing observations by exploiting long-range temporal dependencies, contextual correlations, and behavioral continuity across system states. More importantly, it learns to infer the latent system activities behind incomplete observations, laying the foundation for understanding the underlying dynamics of the cyberspace world.

\subsection{State Chain Prediction}

The state chain prediction task aims to equip the world model with the ability to anticipate future system states based on historical observations. Unlike state chain completion, which reconstructs missing states within an existing trajectory, state chain prediction focuses on forecasting states that have not yet been observed. In dynamic network environments, benign operations and malicious activities evolve over time, and their subsequent behaviors often manifest in changes to code execution, network communications, system execution, and agent interactions. $K$ is the number of future steps after state $S_t$. Learning these evolution patterns enables the world model to anticipate potential system behaviors and identify emerging security risks before they fully materialize.

\begin{equation}
\prod_{j=1}^{K}
P_{\theta}(s_{t+j}\mid
S_{j-1}, C).
\end{equation}

This formulation encourages the model to learn the transition regularities of system states, where each predicted state is conditioned on both historical observations and previously predicted states.

\begin{equation}
-\mathbb{E}{(S,C,R)\sim\mathcal{D}}
\left[
\log P_{\theta}
(S_{t+K},R_{t+K}
\mid S_{0},C)
\right].
\end{equation}

By learning to predict future states and underlying behavioral intentions, the world model captures temporal evolution patterns in running systems. This capability allows StarCore to reason beyond currently observed system states and anticipate subsequent system behaviors, providing a basis for early-stage attack discovery and proactive cybersecurity analysis.

\subsection{State Chain Security Assessment}

The state chain security assessment task aims to enable the world model to understand the security implications of system state evolution and identify potential threats through long-range behavioral reasoning. Unlike conventional security analysis, which classifies isolated events or individual observations, this task requires the model to analyze the complete state chain and investigate relationships among intermediate states. In real-world cyberspace environments, malicious activities may be disguised as legitimate operations at individual time steps, while their underlying attack intentions become evident only through the accumulated behavioral evidence and the interactions among multiple system states. Therefore, effective security assessment requires the world model to trace behavioral transitions, identify abnormal activities, and explain the security implications of the observed system evolution. 

\begin{equation}
P_{\theta}(R\mid S,C,Q)\cdot
P_{\theta}(Y\mid R,S,C).
\end{equation}

Here, $Q$ is the instruction for a complex security task. $R$ represents the intermediate reasoning process for identifying relevant states, tracing their behavioral dependencies, and inferring possible system actions and intentions. $Y$ denotes the final assessment, which summarizes the identified security behaviors, potential attack intentions, and corresponding risks based on the observed evidence.

\begin{equation}
-\mathbb{E}{(S,C,R)\sim\mathcal{D}}
\left[
\log P_{\theta}(R^{},Y^{}
\mid S, C, Q)
\right].
\end{equation}

\begin{figure}[t]
  \centering
  \includegraphics[width=0.95\linewidth]{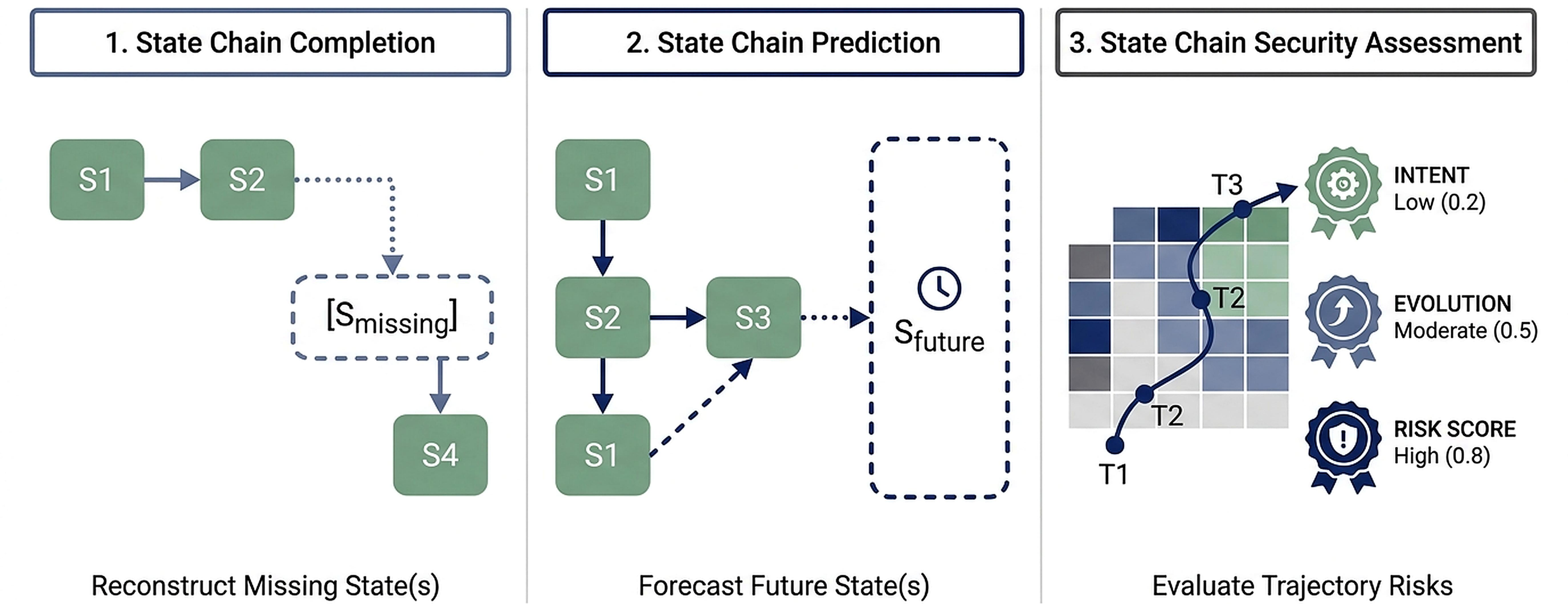}
  \caption{Schematic of the three state-chain supervised fine-tuning (SFT) tasks: (1) State Chain Completion for inferring unobserved intermediate steps; (2) State Chain Prediction for projecting future attack paths; and (3) State Chain Security Assessment for behavioral intent scoring.}
  \label{fig:state_chain_sft_tasks}
\end{figure}

By jointly learning intermediate action reasoning and final security assessments, the world model can connect individual observations into coherent behavioral explanations. This task enables StarCore to move beyond static event classification toward state-chain-based security understanding, supporting attack-behavior tracing, malicious-intention identification, and evidence-based risk assessment in dynamic cyberspace environments. Figure~\ref{fig:state_chain_sft_tasks} illustrates all SFT tasks.

\section{State-Chain Reinforcement Learning}

\begin{figure}[!t]
    \centering
    \includegraphics[width=\linewidth]{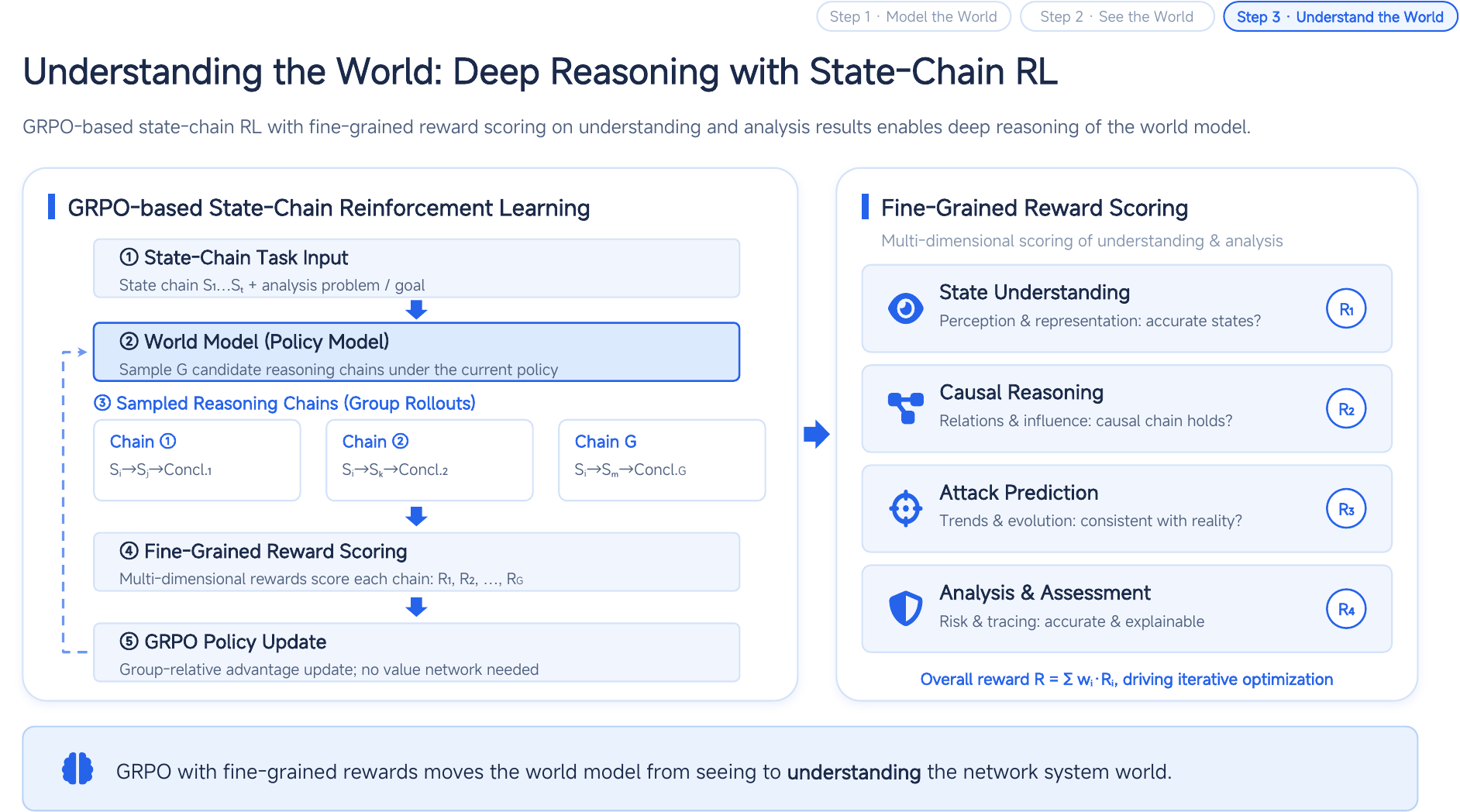}
    \caption{\textbf{Understanding the world:} StarCore employs GRPO to enable deep reasoning with state-chain reinforcement learning, enabling robust reasoning for long-chain analysis.}
    \label{fig:rl}
\end{figure}

The state-chain supervised fine-tuning stage enables StarCore to observe system states, reconstruct incomplete state chains, predict future system behaviors, and perform security analysis. However, supervised learning primarily optimizes the likelihood of predefined reference trajectories, which may limit the model's ability to explore diverse reasoning paths and handle complex, previously unseen system behaviors. To further enhance the world model's understanding of the dynamic cyberspace, we introduce a state-chain reinforcement learning (RL) framework based on Group Relative Policy Optimization (GRPO). Rather than optimizing the textual similarity of generated answers alone, our framework directly evaluates the quality of system state-chain understanding and behavioral reasoning through a multi-dimensional State Chain Reward (SCR).

Specifically, we formulate state-chain reasoning as a reward-driven generation process, where the world model explores multiple candidate reasoning trajectories for each system state-chain input. We design SCR to evaluate generated trajectories across complementary dimensions, including state understanding, causal reasoning, attack prediction, and security assessment, while also considering output-format validity. Based on these fine-grained rewards, we employ GRPO to optimize the policy model through group-relative advantage estimation and clipped policy updates, without introducing an additional value model. This process progressively aligns the world model with the underlying evolution regularities of the cyberspace, improving its ability to conduct long-chain reasoning, understand complex system behaviors, and identify potential security threats.

\subsection{State-Chain Reinforcement Learning Formulation}
\label{sec:rl_formulation}

We formulate state-chain reinforcement learning as a reward-driven conditional generation problem. Given a system state chain, its environmental context, and a task instruction, the world model must generate a reasoning trajectory and a final response that accurately characterize the observed system behaviors. Unlike conventional security-oriented reinforcement learning, which primarily evaluates the correctness of a final classification or textual answer, our formulation treats the intermediate reasoning process as an essential component of the optimization objective.

Formally, we define a state-chain RL input as
\begin{equation}
    x = (\mathcal{I}, Q, S, C, \tau),
\end{equation}
where $\mathcal{I}$ denotes the system instruction, $Q$ represents the task-specific question or analysis objective, $S=\{s_1,s_2,\ldots,s_T\}$ denotes the observed system state chain, $C$ represents the corresponding system context and environmental information, and $\tau\in\mathcal{T}$ indicates the state-chain learning task. Here, $\mathcal{T}=\{\mathrm{comp},\mathrm{pred},\mathrm{assess}\}$ represents state chain completion, prediction, and security assessment, respectively.

The world model is parameterized by a policy $\pi_\theta$, which generates a response trajectory $o=(R, Y)$ conditioned on the input $x$. Specifically, $R$ represents the intermediate reasoning trace that identifies relevant state evidence, infers potential system actions, and explains the behavioral relationships among system states. $Y$ denotes the final task output, including reconstructed states, predicted future states, or security assessment results, depending on the task type. The conditional generation probability is formulated as
\begin{equation}
    \pi_\theta(o\mid x)
    =
    \prod_{t=1}^{|o|}
    \pi_\theta(o_t\mid x,o_{<t}),
\end{equation}
where $o_t$ denotes the $t$-th generated token, $o_{<t}$ represents all preceding generated tokens, and $|o|$ is the response length.

The objective of state-chain RL is to maximize the expected State Chain Reward over the distribution of state-chain inputs:
\begin{equation}
    \max_{\theta}\;
    \mathcal{J}(\theta)
    =
    \mathbb{E}_{x\sim\mathcal{P}_{\mathrm{sc}},
    \,o\sim\pi_\theta(\cdot\mid x)}
    \left[
        R_{\mathrm{SCR}}(x,o)
    \right],
\end{equation}
where $\mathcal{P}_{\mathrm{sc}}$ denotes the distribution of state-chain RL inputs and $R_{\mathrm{SCR}}(x,o)$ evaluates the quality of the generated reasoning trajectory and its corresponding output.

This formulation encourages the world model to discover high-quality reasoning trajectories rather than merely reproduce fixed reference responses. By evaluating both intermediate behavioral reasoning and final task outcomes, state-chain RL provides a direct optimization signal for the model's understanding of system dynamics and security implications.

\subsection{State Chain Reward Design}
\label{sec:scr}

The quality of a cybersecurity world model cannot be sufficiently characterized by the correctness of its final textual answers. In particular, a model may produce a seemingly correct security conclusion while overlooking critical intermediate states, misinterpreting behavioral dependencies, or generating an unsupported attack explanation. To address this limitation, we propose a multi-dimensional State Chain Reward (SCR) that evaluates the model's understanding and reasoning capabilities throughout the state-chain analysis process.

As illustrated in Figure~\ref{fig:rl}, SCR considers five complementary reward components, covering output format validity, state understanding, causal reasoning, attack prediction, and security assessment. Together, these components evaluate the model's ability to perceive system states, understand their evolution, anticipate potential behaviors, and draw reliable security conclusions.

\noindent\textbf{Output Format Reward.}
The output format reward evaluates whether the generated response follows the required structural and serialization constraints. Since state-chain tasks involve heterogeneous system data and structured reasoning results, well-formatted outputs are essential for reliable interpretation and downstream processing. We define the format reward as
\begin{equation}
    r_{\mathrm{fmt}}(x,o)
    =
    \mathcal{V}_{\mathrm{fmt}}(o),
\end{equation}
where $\mathcal{V}_{\mathrm{fmt}}(\cdot)$ is a format verifier that evaluates the structural validity and completeness of the generated response.

\noindent\textbf{State Understanding Reward.}
The state understanding reward measures whether the model correctly identifies and interprets the key information contained in the observed system state chain. It evaluates the model's perception of relevant entities, system behaviors, state attributes, and temporal relationships. In particular, the model should distinguish normal system activities from suspicious observations without overlooking critical evidence. We denote this reward as
\begin{equation}
    r_{\mathrm{state}}(x,o)
    =
    \mathcal{V}_{\mathrm{state}}(S),
\end{equation}
where $\mathcal{V}_{\mathrm{state}}(\cdot)$ evaluates the consistency between the generated state understanding and the available system-state evidence.

\noindent\textbf{Causal Reasoning Reward.}
The causal reasoning reward evaluates whether the model correctly explains the behavioral dependencies and potential actions underlying system state transitions. Instead of simply identifying suspicious states, the model should establish a coherent behavioral trajectory by connecting preceding and succeeding observations and explaining how potential system actions lead to the observed changes. We define
\begin{equation}
    r_{\mathrm{causal}}(x,o)
    =
    \mathcal{V}_{\mathrm{causal}}(S,R),
\end{equation}
where $\mathcal{V}_{\mathrm{causal}}(\cdot)$ evaluates the logical consistency of the inferred behavioral relationships, the supporting state evidence, and the corresponding action reasoning. This reward encourages the model to develop coherent explanations of system evolution rather than generate isolated or unsupported behavioral descriptions.

\noindent\textbf{Attack Prediction Reward.}
The attack prediction reward measures the model's ability to anticipate future system behaviors and identify potential attack developments from historical state chains. It evaluates whether the predicted states and behavioral trends match the actual system evolution, particularly for emerging or multi-stage malicious activities. We formulate this reward as
\begin{equation}
    r_{\mathrm{pred}}(x,o)
    =
    \mathcal{V}_{\mathrm{pred}}
    (S_{\mathrm{future}},R,Y),
\end{equation}
where $S_{\mathrm{future}}$ denotes the reference future system states when available, and $\mathcal{V}_{\mathrm{pred}}(\cdot)$ evaluates the consistency of the generated predictions and their underlying behavioral reasoning with the reference evolution.

\noindent\textbf{Security Assessment Reward.}
The security assessment reward evaluates the correctness and evidential support of the final security analysis. It considers whether the model accurately identifies abnormal behaviors, infers potential malicious intentions, and derives security conclusions consistent with the observed state-chain evidence. Importantly, the model should avoid unsupported attack attribution and distinguish legitimate system activity from malicious behavior. We define
\begin{equation}
    r_{\mathrm{assess}}(x,o)
    =
    \mathcal{V}_{\mathrm{assess}}(S,R,Y),
\end{equation}
where $\mathcal{V}_{\mathrm{assess}}(\cdot)$ evaluates the correctness, completeness, and evidence consistency of the generated security assessment.

The overall SCR is computed as a weighted aggregation of these reward components:
\begin{equation}
\begin{aligned}
    R_{\mathrm{SCR}}(x,o)
    ={}&
    w_{\mathrm{fmt}}r_{\mathrm{fmt}}
    +w_{\mathrm{state}}r_{\mathrm{state}}\\
    &+w_{\mathrm{causal}}r_{\mathrm{causal}}
    +w_{\mathrm{pred}}r_{\mathrm{pred}}\\
    &+w_{\mathrm{assess}}r_{\mathrm{assess}},
\end{aligned}
\label{eq:scr}
\end{equation}
where each reward component is normalized to a comparable range, and $w_{\mathrm{fmt}}$, $w_{\mathrm{state}}$, $w_{\mathrm{causal}}$, $w_{\mathrm{pred}}$, and $w_{\mathrm{assess}}$ are non-negative weights controlling their respective contributions. The weights can be configured according to the relative importance of each evaluation dimension.

Because different state-chain tasks emphasize different capabilities, the reward components activate according to the task type $\tau$. For example, completion emphasizes the correctness of reconstructed states, prediction emphasizes the consistency of future-state forecasts, and security assessment emphasizes behavioral reasoning and security conclusions. Accordingly, the general reward formulation can be expressed as
\begin{equation}
    R_{\mathrm{SCR}}^{(\tau)}(x,o)
    =
    \sum_{k\in\mathcal{K}}
    m_{\tau,k}w_k r_k(x,o),
\end{equation}
where $\mathcal{K}$ denotes the set of reward dimensions, and $m_{\tau,k}\in\{0,1\}$ indicates whether the $k$-th reward component is applicable to task $\tau$. The weights of the active components are normalized such that
\begin{equation}
    \sum_{k\in\mathcal{K}}m_{\tau,k}w_k=1.
\end{equation}

Through this multi-dimensional reward design, SCR evaluates not only whether the model produces a correct answer, but also whether it understands the observed system states, establishes reasonable behavioral relationships, and provides evidence-supported security reasoning. This lets the reinforcement learning process optimize underlying world-understanding capabilities rather than merely improving surface-level answer generation.

\subsection{GRPO-based Policy Optimization}
\label{sec:grpo}

Based on the proposed SCR, we employ Group Relative Policy Optimization (GRPO) to optimize the state-chain reasoning policy. GRPO is particularly suitable for our framework because it enables the model to explore multiple reasoning trajectories for the same system state-chain input and directly compare their relative quality using group-level rewards, without requiring an additional value network.

Specifically, given a state-chain input $x$, the current policy $\pi_{\theta_{\mathrm{old}}}$ independently samples a group of $G$ candidate response trajectories:
\begin{equation}
    \mathcal{O}(x)
    =
    \{o_1,o_2,\ldots,o_G\},
    \qquad
    o_i\sim\pi_{\theta_{\mathrm{old}}}(\cdot\mid x),
\end{equation}
where $\theta_{\mathrm{old}}$ denotes the parameters of the policy used to generate the current rollout group, and $G$ is the number of sampled candidates. Each candidate consists of an intermediate reasoning trace and a final task response. We then evaluate each candidate using SCR to obtain its scalar reward:
\begin{equation}
    r_i=R_{\mathrm{SCR}}(x,o_i),
    \qquad i=1,\ldots,G.
\end{equation}

To estimate the relative quality of each candidate, GRPO normalizes the rewards within the sampled group. Specifically, the group mean and standard deviation are calculated as
\begin{equation}
    \mu_r
    =
    \frac{1}{G}\sum_{i=1}^{G}r_i,
\end{equation}
\begin{equation}
    \sigma_r
    =
    \sqrt{
    \frac{1}{G}
    \sum_{i=1}^{G}(r_i-\mu_r)^2
    }.
\end{equation}
The group-relative advantage of the $i$-th response is subsequently defined as
\begin{equation}
    \hat{A}_i
    =
    \frac{r_i-\mu_r}{\sigma_r+\epsilon},
\label{eq:grpo_advantage}
\end{equation}
where $\epsilon$ is a small constant for numerical stability. A positive advantage indicates that a candidate performs better than the group average, whereas a negative advantage indicates relatively inferior performance. This relative comparison encourages the policy to increase the probability of higher-quality reasoning trajectories while reducing the probability of inferior ones.

To optimize the policy while preventing excessively large updates, we use the clipped surrogate objective from GRPO. For the $t$-th token of the $i$-th sampled response, the probability ratio between the updated policy and the rollout policy is defined as
\begin{equation}
    \rho_{i,t}(\theta)
    =
    \frac{
    \pi_\theta(o_{i,t}\mid x,o_{i,<t})
    }{
    \pi_{\theta_{\mathrm{old}}}(o_{i,t}\mid x,o_{i,<t})
    }.
\end{equation}
The clipped policy optimization objective is formulated as
\begin{equation}
\begin{aligned}
    \mathcal{J}_{\mathrm{clip}}(\theta)
    =
    \mathbb{E}_{x,\mathcal{O}(x)}
    \left[
    \frac{1}{G}
    \sum_{i=1}^{G}
    \frac{1}{|o_i|}
    \sum_{t=1}^{|o_i|}
    \min\left(
        \rho_{i,t}(\theta)\hat{A}_i,\right.\right.\\
    \left.\left.
        \operatorname{clip}
        \left(
            \rho_{i,t}(\theta),
            1-\varepsilon,
            1+\varepsilon
        \right)\hat{A}_i
    \right)
    \right],
\end{aligned}
\label{eq:grpo_clip}
\end{equation}
where $\varepsilon$ is the clipping parameter that constrains the magnitude of policy updates, and $\operatorname{clip}(\cdot)$ restricts the probability ratio to the interval $[1-\varepsilon,1+\varepsilon]$. The sequence-level advantage $\hat{A}_i$ is shared across all generated tokens of the corresponding response, allowing the policy to optimize the overall quality of each reasoning trajectory.

Furthermore, to preserve the security knowledge and state-chain understanding acquired during the preceding training stages, we introduce a Kullback-Leibler (KL) divergence constraint between the updated policy and a reference policy $\pi_{\mathrm{ref}}$, which is initialized from the state-chain SFT model and remains frozen during RL optimization. The final GRPO objective is given by
\begin{equation}
\begin{aligned}
    \mathcal{J}_{\mathrm{GRPO}}(\theta)
    =
    \mathcal{J}_{\mathrm{clip}}(\theta)
    -
    \beta
    \mathbb{E}_{x,o\sim\pi_\theta}
    \left[
        D_{\mathrm{KL}}
        \left(
            \pi_\theta(\cdot\mid x)
            \,\|\,
            \pi_{\mathrm{ref}}(\cdot\mid x)
        \right)
    \right],
\end{aligned}
\label{eq:grpo_objective}
\end{equation}
where $\beta$ controls the strength of the KL regularization. This constraint prevents the optimized policy from deviating excessively from the reference model, retaining its acquired cybersecurity knowledge and state-chain understanding while improving reasoning quality.

The GRPO optimization proceeds iteratively. At each iteration, the policy generates multiple candidate reasoning trajectories for each state-chain input, evaluates them using SCR, computes group-relative advantages, and updates the policy by maximizing the clipped objective with KL regularization. Unlike conventional actor--critic methods, this process does not require an additional value model to estimate token-level or state-level values, reducing the additional parameters and training overhead.

Overall, the proposed state-chain reinforcement learning framework combines multi-dimensional behavioral evaluation with group-relative policy optimization. By explicitly rewarding accurate state understanding, coherent behavioral reasoning, reliable attack prediction, and evidence-based security assessment, StarCore progressively improves its ability to understand how complex system behaviors evolve. This enables the model to move beyond merely observing and reproducing system state chains toward deeply reasoning about the dynamic cyberspace and underlying security implications.

\section{Evaluation}
\label{sec:evaluation}

\subsection{Experimental Setup}

\noindent \textbf{Hardware Environment.}
We develop and evaluate StarCore on two GPU platforms, NVIDIA B300 and A800. 
These platforms provide the computational resources for model development, training, and inference evaluation. 
We run experiments on both platforms to support development and testing of the proposed cybersecurity world model under different hardware configurations.

\noindent \textbf{Dataset Collection.}
To build a comprehensive corpus for cybersecurity world modeling, we collect publicly available data from multiple open-source platforms and research resources, including Hugging Face, GitHub, and ModelScope, as well as open-source repositories and datasets released through cybersecurity research. 
Rather than relying on a single data source or format, our collection covers diverse artifacts generated by real-world computing systems and security-related activities.

\begin{figure}[t]
  \centering
  \includegraphics[width=0.95\linewidth]{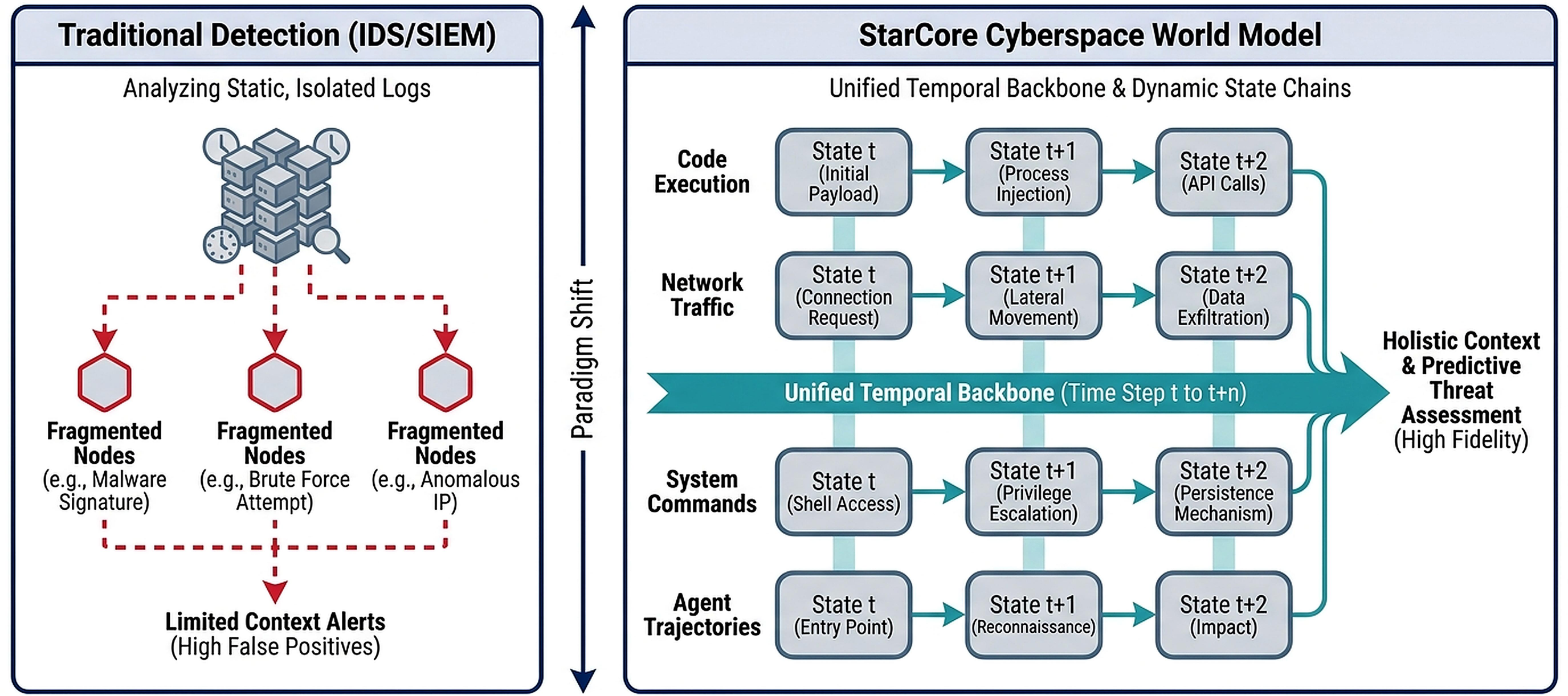}
  \caption{Comparison between traditional cybersecurity threat detection and the world model paradigm. Unlike static tools analyzing isolated logs, StarCore continuously perceives dynamic state chains spanning code execution, network traffic, system commands, and agent trajectories over unified time horizons.}
  \label{fig:cyber_world_model_overview}
\end{figure}

\noindent \textbf{Data Modalities.}
The collected corpus spans four major modalities: (1) \emph{source code}, including program implementations and code-related artifacts; (2) \emph{network traffic}, including packet- and flow-level data; (3) \emph{operating system logs}, including system events and execution records; and (4) \emph{agent trajectories}, including interaction histories and action-related records produced by intelligent agents. 
These modalities capture complementary aspects of system behavior, ranging from program logic and network communication to runtime events and agent interactions. 
By integrating heterogeneous data sources, StarCore is designed to learn system states and their temporal and causal dependencies across different cybersecurity scenarios. Figure~\ref{fig:cyber_world_model_overview} shows the world model's data-chain paradigm compared with traditional models.

\noindent \textbf{Data Preparation.}
Before training and evaluation, we organize the collected data into structured state-chain samples following the data format described in Section~3.1. 
Each sample includes task instructions and corresponding questions, along with the available system states, action reasoning, and expected answers. 
This unified representation enables heterogeneous cybersecurity artifacts to be incorporated into the state-chain learning tasks, including state-chain completion, state-chain prediction, and state-chain security assessment.

\subsection{Overall Performance Evaluation}

\begin{table*}[t]
\centering
\caption{Answer accuracy comparison of different LLMs across system data modalities and system-world understanding tasks. Overall accuracy is calculated as the arithmetic mean of the seven evaluation dimensions. The best and second-best results in each column are highlighted in bold and underlined, respectively.}
\label{tab:overall_performance}
\setlength{\tabcolsep}{5.5pt}
\renewcommand{\arraystretch}{1.18}
\resizebox{\textwidth}{!}{
\begin{tabular}{lcccccccc}
\toprule
\multirow{2}{*}{\textbf{Model}}
& \multicolumn{4}{c}{\textbf{System Data Types}}
& \multicolumn{3}{c}{\textbf{System-World Understanding Tasks}}
& \multirow{2}{*}{\textbf{Overall}} \\
\cmidrule(lr){2-5}
\cmidrule(lr){6-8}
& \textbf{Code}
& \textbf{Agent}
& \textbf{Traffic}
& \textbf{Command}
& \textbf{Security}
& \textbf{Completion}
& \textbf{Prediction}
& \\
\midrule
Qwen3.5-9B
& 0.7496 & 0.7525 & 0.7724 & 0.6508
& 0.8845 & 0.6426 & 0.6596 & 0.7328 \\

Qwen3.5-35B
& 0.7325 & 0.8279 & 0.7931 & 0.6669
& 0.9012 & 0.6709 & 0.6878 & 0.7561 \\

DeepSeek-v4-Flash
& 0.5771 & 0.6345 & 0.0945 & 0.2345
& 0.4776 & 0.3796 & 0.2962 & 0.3847 \\

DeepSeek-v4-Pro
& 0.5632 & 0.6860 & 0.5636 & 0.4112
& 0.5343 & 0.5398 & 0.5946 & 0.5563 \\

MiniMax M3
& 0.7550 & 0.7636 & 0.7071 & 0.6337
& 0.8771 & 0.6201 & 0.6584 & 0.7134 \\

Kimi K3
& 0.7344 & 0.5272 & 0.5496 & 0.5539
& 0.6765 & 0.5632 & 0.5170 & 0.5902 \\

GLM-5.2
& 0.7175 & 0.7980 & 0.6436 & 0.5165
& 0.7731 & 0.6258 & 0.6069 & 0.6688 \\

GLM-5.3
& 0.6329 & 0.5696 & 0.6084 & 0.5569
& 0.8269 & 0.4652 & 0.4820 & 0.5919 \\

Qwen3-235B-A22B
& 0.7931 & 0.5049 & 0.6604 & 0.3860
& 0.6000 & 0.6714 & 0.5387 & 0.6090 \\

Qwen3.8-Flash-Next 
& 0.6950 & 0.7472 & 0.5957 & 0.6440
& 0.8176 & 0.6051 & 0.5893 & 0.6705 \\

GLM-5.3-Flash
& 0.6980 & 0.6273 & 0.6851 & 0.6018
& 0.8358 & 0.5318 & 0.5901 & 0.6531 \\

\midrule
\rowcolor{green!12}
\textsf{StarCore-9B}
& 0.7564 & 0.9260 & 0.8678 & 0.7297
& 0.9040 & 0.8380 & 0.7207 & 0.8201 \\

\rowcolor{green!12}
\textsf{StarCore-35B}
& \underline{0.8929} & 0.9104 & \underline{0.9339} & \underline{0.8117}
& 0.8716 & \underline{0.9122} & \underline{0.8781} & \underline{0.8872} \\

\rowcolor{green!20}
\textbf{\textsf{StarCore-180B}}
& \textbf{0.9306} & \textbf{0.9603} & \textbf{0.9672} & \textbf{0.8371}
& \textbf{0.9429} & \textbf{0.9301} & \textbf{0.8983}
& \textbf{0.9238} \\

\rowcolor{green!20}
\textsf{StarCore-320B}
& 0.8904 & \underline{0.9470} & 0.9100 & 0.7738
& \underline{0.9313} & \underline{0.9237} & 0.7851
& 0.8802 \\

\bottomrule
\end{tabular}
}
\end{table*}

We first evaluate StarCore's answer accuracy by comparing it with a diverse set of state-of-the-art large language models, including Qwen\cite{qwen3}, DeepSeek\cite{deepseek2026v4}, MiniMax\cite{minimax2026m3}, Kimi\cite{kimi-k3}, and GLM\cite{glm-5}. The evaluation covers four system data modalities, namely code, agent trajectories, network traffic, and Linux commands, as well as three core system-world understanding tasks: security analysis, state completion, and state prediction. For each evaluation dimension, answer accuracy measures the proportion of model-generated answers that are correct. Table~\ref{tab:overall_performance} reports answer accuracy across these seven dimensions, along with overall accuracy.

\noindent \textbf{Overall Answer Accuracy.}
As shown in Table~\ref{tab:overall_performance}, StarCore achieves consistently high answer accuracy across different model scales. In particular, StarCore-180B obtains the highest overall accuracy of 0.9238, outperforming the strongest baseline, Qwen3.5-35B (0.7561), by 0.1677. This result demonstrates the effectiveness of state-chain learning on interpreting varied system behaviors and reasoning about their security implications. Moreover, the accuracy improvements span all seven evaluation dimensions, indicating that the benefits of StarCore are not confined to a particular data modality or task.

\noindent \textbf{Answer Accuracy across System Data Modalities.}
We further examine answer accuracy across the four system data modalities. StarCore-180B achieves accuracies of 0.9306, 0.9603, 0.9672, and 0.8371 on code, agent trajectories, network traffic, and Linux commands, respectively, outperforming all baseline models in each dimension. Compared with the strongest baseline in each category, these results represent improvements of 13.75, 13.24, 17.41, and 17.02\%, respectively. In particular, the high accuracy on network traffic and Linux commands highlights the model's ability to interpret operating system behavior beyond conventional code understanding. These results show that StarCore can produce reliable answers across different system data modalities, not limited to a single type of system data.

\noindent \textbf{Answer Accuracy on System-World Understanding Tasks.}
Beyond modality-specific understanding, we evaluate whether models can correctly answer questions involving higher-level system-world reasoning through security analysis, state completion, and state prediction. StarCore-180B achieves accuracies of 0.9429, 0.9301, and 0.8983 on these three tasks, respectively, consistently outperforming all baseline models. In particular, it improves answer accuracy on state completion and prediction by 25.87 and 21.05\%, respectively, over the strongest baseline on each task. These substantial improvements highlight the benefits of learning system-state dependencies and temporal evolution. Rather than merely interpreting individual system artifacts, StarCore can leverage contextual information to correctly reconstruct missing states and predict subsequent system behaviors. Meanwhile, its 0.9429 accuracy on security analysis demonstrates its strong ability to interpret system behavior and produce accurate security assessments.

\noindent \textbf{Answer Accuracy across Model Scales.}
We also evaluate StarCore at different model scales, including 9B, 35B, 180B, and 320B parameters. Even at the 9B scale, StarCore-9B achieves an overall accuracy of 0.8201, outperforming Qwen3.5-9B (0.7328) by 8.73\%. At the 35B scale, StarCore-35B further achieves an overall accuracy of 0.8872, substantially outperforming Qwen3.5-35B (0.7561). The 180B variant achieves the best overall accuracy of 0.9238, indicating that the proposed state-chain learning framework delivers substantial improvements across model scales. Although the 320B variant remains competitive with an overall accuracy of 0.8802, it is lower than the 180B variant. This observation suggests that increasing model size alone does not necessarily lead to higher answer accuracy. Overall, these results show that StarCore's improvements are not solely attributable to parameter scaling, but also reflect the benefits of explicitly learning system-state dependencies and dynamics.

\begin{table}[t]
\centering
\caption{\label{tab:reasoning_comparison} Comparison of reasoning correctness across different LLMs on the BigModelGym benchmark. Higher values indicate more accurate reasoning rationales. The best and second-best results in each row are highlighted in bold and underlined, respectively.}
\setlength{\tabcolsep}{5pt}
\renewcommand{\arraystretch}{1.15}
\resizebox{\columnwidth}{!}{
\begin{tabular}{lcccccccc}
\toprule
& \textbf{StarCore-9B}
& \textbf{MiniMax M3}
& \textbf{GLM-5.2}
& \textbf{Qwen3.5-35B}
& \textbf{Qwen3.5-9B}
& \begin{tabular}[c]{@{}c@{}}\textbf{DeepSeek}\\\textbf{v4-Flash}\end{tabular}
& \begin{tabular}[c]{@{}c@{}}\textbf{DeepSeek}\\\textbf{v4-Pro}\end{tabular}
& \textbf{Kimi K3} \\
\midrule

\textit{Overall} 
& \textbf{52.8} & \underline{47.7} & 44.0 & 43.9 & 42.7 & 40.2 & 39.8 & 39.6 \\

\midrule
\textit{System-World Tasks} & & & & & & & & \\

Security Tasks
& \textbf{57.0} & 48.0 & \underline{49.8} & 47.9 & 49.1 & 35.9 & 44.4 & 35.7 \\

Completion Tasks
& \textbf{60.2} & \underline{55.7} & 44.2 & 50.9 & 47.7 & 38.4 & 43.0 & 44.6 \\

Prediction Tasks
& 41.1 & 38.4 & 38.0 & 32.5 & 31.1 & \textbf{46.3} & 32.0 & 36.6 \\

\midrule
\textit{System Data Types} & & & & & & & & \\

Code Data
& \textbf{76.4} & \underline{71.0} & 62.5 & 63.7 & 61.3 & 55.4 & 56.5 & 55.5 \\

Agent Trajectory
& \textbf{42.2} & 34.3 & 35.9 & 39.2 & 37.9 & 38.4 & 33.1 & \underline{38.7} \\

Traffic Data
& \textbf{66.6} & \underline{57.5} & 55.3 & 52.2 & 52.4 & 28.6 & 52.6 & 45.4 \\

System Command 
& 25.9 & 26.8 & 22.3 & 21.2 & 18.2 & \textbf{38.4} & 16.8 & 20.3 \\

\bottomrule
\end{tabular}
}
\end{table}

\subsection{Reasoning Capability Evaluation}
In addition to evaluating answer accuracy, we further investigate whether StarCore can produce correct and logically grounded reasoning processes when understanding complex system behaviors. To this end, we independently evaluate reasoning correctness using the StarCore benchmark on BigModelGym~\cite{bigmodelgym}. Unlike answer evaluation, which measures the correctness of final answers, this evaluation independently assesses whether the reasoning rationales generated by models are correct. As described by the benchmark, each evaluation dimension contains 250 samples or corresponding sample combinations, with 1,000 samples in each leaderboard. The reasoning evaluation covers four system data types (code, agent trajectories, network traffic, and system commands) and three system-world understanding tasks (security analysis, state completion, and state prediction). The benchmark calculates the overall score as a weighted aggregation across eight tasks.

As shown in Table~\ref{tab:reasoning_comparison}, StarCore-9B achieves the highest overall reasoning score of 52.8, outperforming the second-best model, MiniMax M3 (47.7), by 5.1\%. Notably, StarCore-9B achieves the best results in five evaluation dimensions, including security analysis (57.0), state completion (60.2), code understanding (76.4), agent trajectory understanding (42.2), and network traffic understanding (66.6). These results demonstrate its strong ability to reason about heterogeneous system artifacts and reconstruct the underlying system behaviors. In particular, its leading performance on state completion suggests that the model can effectively leverage contextual evidence to infer missing system states, while its performance on code and network traffic indicates its ability to establish meaningful connections between different types of system evidence.

We also observe that StarCore-9B does not achieve the highest score in every individual dimension. For example, DeepSeek-v4-Flash performs better on state prediction (46.3) and system command understanding (38.4). Nevertheless, StarCore-9B maintains the highest overall reasoning score, indicating a more balanced reasoning capability across heterogeneous system data and understanding tasks. This is particularly important for cybersecurity world models, where reliable reasoning requires integrating diverse system evidence rather than excelling at an isolated task.

Taken together with the answer accuracy evaluation in Table~\ref{tab:overall_performance}, we find that StarCore improves both final-answer correctness and reasoning correctness. The answer accuracy evaluation measures whether the model reaches the correct conclusion, whereas the reasoning evaluation examines whether the rationale supporting that conclusion is itself correct. Together, these results show that state-chain learning improves not only output correctness but also the quality of the underlying system-world reasoning process.

\subsection{Model Scale Analysis}
\label{subsec:model-scale-analysis}

\begin{figure}[t]
    \centering
    \includegraphics[width=0.98\textwidth]{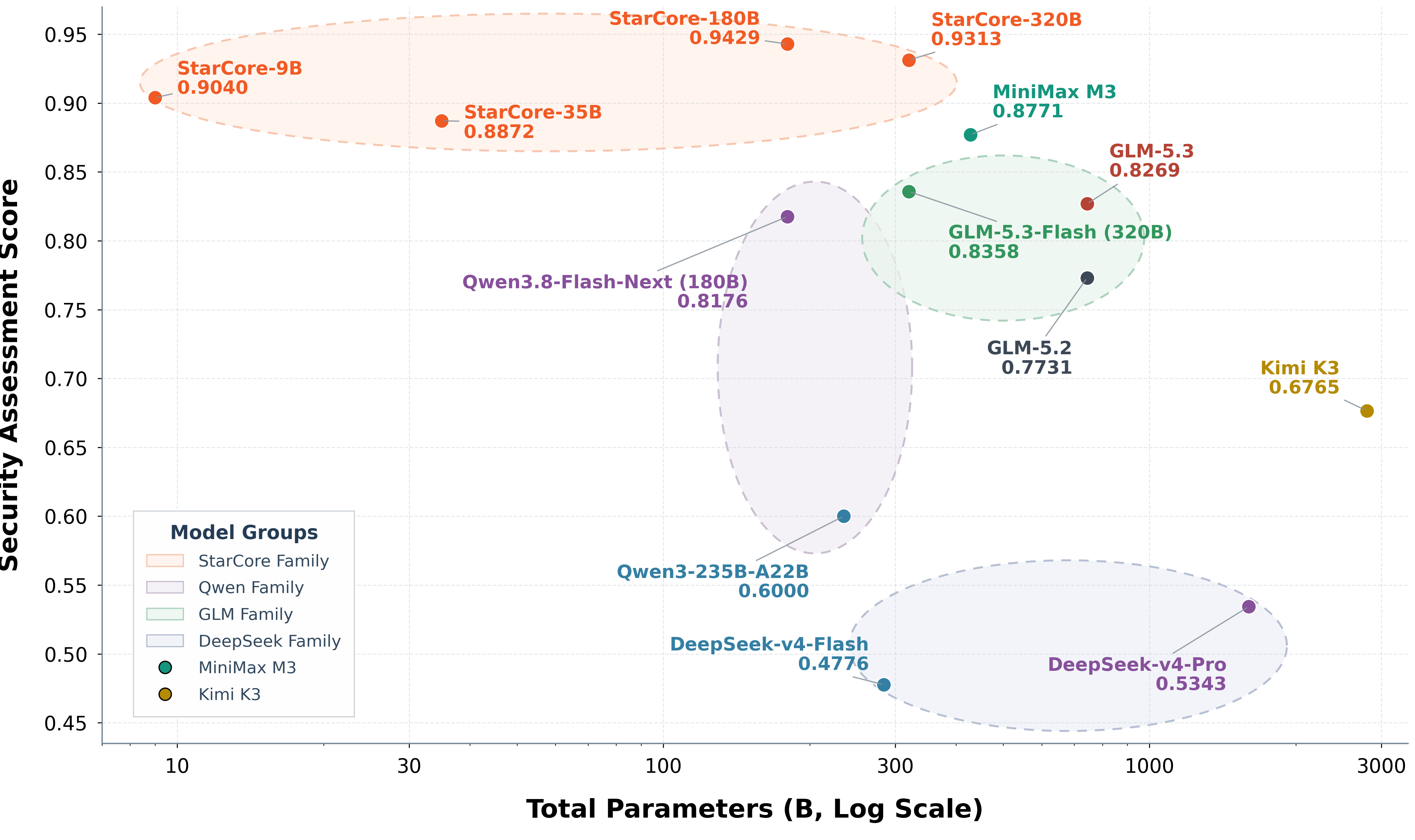}
    \caption{Relationship between total parameter count and security assessment accuracy. The horizontal axis uses a logarithmic scale. Shaded ellipses denote model families, while MiniMax M3 and Kimi K3 are shown as individual models.}
    \label{fig:model-scale-security}
\end{figure}

To further investigate whether stronger security assessment capability primarily arises from parameter scaling, we analyze the relationship between total parameter count and security assessment accuracy. As shown in Figure~\ref{fig:model-scale-security}, the horizontal axis represents the total number of model parameters on a logarithmic scale, while the vertical axis reports accuracy on the security assessment task. Shaded ellipses group models from the same family, enabling a joint examination of scaling trends within individual families and differences across families. This analysis complements the overall results in Table~1 by focusing specifically on security assessment, which requires models to interpret system-state evidence, reason about behavioral dependencies, and derive reliable security conclusions.

\textbf{Model Scale Alone Does Not Determine Security Assessment Accuracy.} The results reveal no consistent monotonic relationship between parameter count and security assessment accuracy across the evaluated models. For example, Qwen3.8-Flash-Next achieves an accuracy of 0.8176 at the 180B scale, whereas Qwen3-235B-A22B obtains 0.6000 despite having a larger total parameter count. Similarly, Kimi K3, with 2,800B parameters, achieves 0.6765, still below several substantially smaller models. Within the DeepSeek family, increasing the model scale from DeepSeek-v4-Flash to DeepSeek-v4-Pro improves accuracy from 0.4776 to 0.5343, but the larger model still does not reach the performance of many smaller counterparts. These observations indicate that parameter scaling alone is insufficient to guarantee accurate interpretation of security-relevant system behaviors.

\textbf{Intra-Family Differences Highlight the Role of Model Design and Training.} Models within the same family also exhibit substantial variation at comparable or even identical scales. GLM-5.2 and GLM-5.3, both at approximately 744B total parameters in our analysis, achieve security assessment accuracies of 0.7731 and 0.8269, respectively, an absolute difference of 0.0538. Moreover, GLM-5.3-Flash achieves 0.8358 at the smaller 320B scale, exceeding both larger variants. The Qwen family shows an even more pronounced difference: the 180B Qwen3.8-Flash-Next outperforms Qwen3-235B-A22B by 0.2176 in absolute accuracy. These within-family differences suggest that backbone design, training data, post-training strategies, and alignment with security reasoning tasks can substantially affect performance beyond parameter count.

\textbf{StarCore Demonstrates Strong Accuracy--Scale Trade-offs.} All four StarCore variants outperform the strongest non-StarCore model, MiniMax M3, which achieves an accuracy of 0.8771 at the 428B scale. Notably, StarCore-9B obtains 0.9040, exceeding MiniMax M3 by 0.0269 despite being approximately 47 times smaller in total parameter count. StarCore-35B also remains ahead of MiniMax M3 with an accuracy of 0.8872. These results demonstrate that state-chain learning can provide strong security assessment capability without relying on extremely large model scales. Although our evaluation does not directly measure inference latency or computational cost, the performance of the compact variants suggests a favorable accuracy--scale trade-off for deployment scenarios with limited computational resources.

\textbf{StarCore Exhibits Non-Monotonic Scaling Behavior.} Among the evaluated models, StarCore-180B achieves the highest security assessment accuracy of 0.9429, improving upon StarCore-9B by 0.0389. StarCore-320B also performs strongly with an accuracy of 0.9313, but remains 0.0116 below StarCore-180B. In addition, StarCore-35B does not surpass the 9B variant on this task. Therefore, the benefits of StarCore should not be interpreted as a simple consequence of increasing model size. Instead, the results suggest that state-chain supervised fine-tuning and reinforcement learning play a central role in establishing accurate security reasoning, while the final performance at each scale may also depend on the backbone, training configuration, and task alignment.

Overall, Figure~\ref{fig:model-scale-security} shows that explicit modeling of system-state dependencies can be more important for security assessment than parameter scaling alone. The strong performance of StarCore-9B and StarCore-35B supports the feasibility of compact cybersecurity world models, whereas StarCore-180B provides the highest accuracy among the evaluated variants. Nevertheless, these observations are specific to the security assessment task and should not be regarded as a universal scaling law~\cite{law}. Together with the overall accuracy results in Table~1 and the reasoning evaluation in Table~2, they indicate that reliable cybersecurity world modeling requires both appropriate model capacity and explicit learning of system-world states, transitions, and security implications.
\section{Capability Analysis and Case Study}
\label{subsec:code-security-capability}

To examine whether the system-world understanding capabilities of StarCore transfer to code-level security analysis, we conduct an exploratory study on a mixed collection of code snippets. The original collection contains 83 samples with reference answers but no independently verified vulnerability labels. We deterministically select 50 samples using a fixed hash-based rule, resulting in 13 reference-positive and 37 reference-negative cases. Five models are evaluated under two conditions: (1) \emph{binary classification}, which requests only a vulnerable-or-benign decision, and (2) \emph{evidence explanation}, which additionally requests the vulnerability mechanism, code evidence, assumptions, and limitations. The code and question are identical in both conditions. All 500 expected model-prompt-sample combinations are recorded, and failed responses are retained in the availability analysis rather than replaced through selective rerunning.

We compute precision, recall, and F1 over valid responses using the provided reference labels, whereas we report response availability over all 50 samples. Consequently, these metrics measure agreement with the supplied references rather than verified performance on an independent vulnerability ground truth. This distinction is essential because the reference collection contains both directly inspectable risks and cases whose labels depend on missing caller, memory-safety, or runtime context.

\begin{figure}[t]
    \centering
    \includegraphics[width=0.98\textwidth]{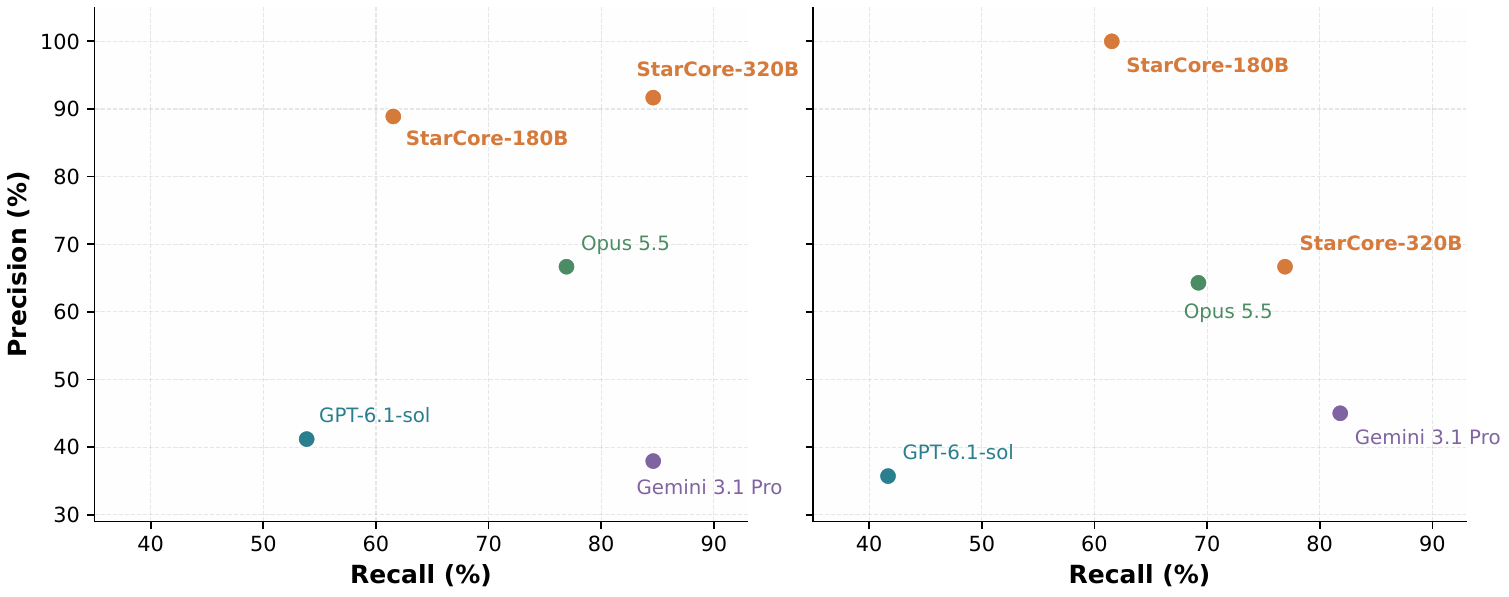}
        \begin{minipage}[t]{0.48\textwidth}
        \centering
        \subcaption{Judgment with binary classification.}
        \label{fig:code-security-pr-binary}
    \end{minipage}
    \hfill
    \begin{minipage}[t]{0.48\textwidth}
        \centering
        \subcaption{Judgment with evidence explanation.}
        \label{fig:code-security-pr-explanation}
    \end{minipage}
    \caption{Precision--recall comparison on the 50-sample code security set under (a) judgment with binary classification and (b) judgment with evidence explanation.}
    \label{fig:code-security-pr}
\end{figure}

\textbf{Judgment with Binary Classification.} Figure~\ref{fig:code-security-pr}(a) shows that StarCore-320B provides the strongest binary classification result under the reference-label protocol. It detects 11 of the 13 reference-positive cases with only one reference false positive, yielding 91.67\% precision, 84.62\% recall, and an F1 score of 88.00\%. StarCore-180B exhibits a more conservative operating point: its precision reaches 88.89\%, but its recall is 61.54\%, corresponding to eight detected reference-positive cases and one reference false positive. These two variants therefore illustrate different detection behaviors rather than a single monotonic scaling trend.

The baseline models occupy distinct regions of the precision--recall space. As shown in Table~\ref{tab:binary-vulnerability-classification}, Gemini 3.1 Pro~\cite{google2026gemini31pro} also detects 11 reference-positive cases, but produces 18 reference false positives and achieves only 37.93\% precision. Opus 5.5~\cite{anthropic2026opus55} provides a more balanced result, with 66.67\% precision and 76.92\% recall, whereas GPT-6.1-sol~\cite{openai2026gpt61sol} obtains 41.18\% precision and 53.85\% recall. Overall, the binary condition demonstrates that StarCore can attain high agreement with the supplied references, particularly at the 320B scale. However, reference agreement alone does not establish whether the model has identified the correct vulnerability mechanism or whether the reference label itself is valid.

\begin{table}[t!]
    \centering
    \caption{Code security judgment results on the 50-sample code security set. Precision, recall, and F1 are computed over valid responses, while full-set agreement is calculated over all 50 samples.}
    \label{tab:binary-vulnerability-classification}
    \small
    \begin{tabular}{@{}lcccccc@{}}
        \toprule
        Model & Valid Responses & Precision & Recall & F1 & Full-Set Agreement & TP/TN/FP/FN \\
        \midrule
        StarCore-180B  & 50/50 & 88.89\% & 61.54\% & 72.73\% & 88.00\% & 8/36/1/5 \\
        StarCore-320B  & 50/50 & \textbf{91.67\%} & \textbf{84.62\%} & \textbf{88.00\%} & \textbf{94.00\%} & 11/36/1/2 \\
        GPT-6.1-sol    & 50/50 & 41.18\% & 53.85\% & 46.67\% & 68.00\% & 7/27/10/6 \\
        Gemini 3.1 Pro & 49/50 & 37.93\% & 84.62\% & 52.38\% & 58.00\% & 11/18/18/2 \\
        Opus 5.5       & 50/50 & 66.67\% & 76.92\% & 71.43\% & 84.00\% & 10/32/5/3 \\
        \bottomrule
    \end{tabular}
\end{table}

\begin{table}[t!]
    \centering
    \caption{Code vulnerability classification results under judgment with evidence explanation.}
    \label{tab:explanation-vulnerability-classification}
    \small
    \begin{tabular}{@{}lcccccc@{}}
        \toprule
        Model & Valid Responses & Precision & Recall & F1 & Full-set Agreement & TP/TN/FP/FN \\
        \midrule
        StarCore-180B  & 50/50 & \textbf{100.00\%} & 61.54\% & \textbf{76.19\%} & \textbf{90.00\%} & 8/37/0/5 \\
        StarCore-320B  & 50/50 & 66.67\%  & \textbf{76.92\%} & 71.43\% & 84.00\% & 10/32/5/3 \\
        GPT-6.1-sol    & 48/50 & 35.71\%  & 41.67\% & 38.46\% & 64.00\% & 5/27/9/7 \\
        Gemini 3.1 Pro & 41/50 & 45.00\%  & 81.82\% & 58.06\% & 56.00\% & 9/19/11/2 \\
        Opus 5.5       & 49/50 & 64.29\%  & 69.23\% & 66.67\% & 80.00\% & 9/31/5/4 \\
        \bottomrule
    \end{tabular}
\end{table}

\textbf{Judgment with Evidence Explanation.} As shown in Figure~\ref{fig:code-security-pr}(b), StarCore-180B increases its precision to 100\% while leaving recall unchanged at 61.54\%. The paired responses show that this change results from correcting one reference false positive, without detecting any additional reference-positive case. In contrast, GPT-6.1-sol and Opus 5.5 show mixed paired transitions rather than consistent gains.

As shown in Table~\ref{tab:explanation-vulnerability-classification}, the two StarCore variants respond substantially differently to this prompt condition. The baseline models likewise show no consistent benefit from explanation prompting. GPT-6.1-sol obtains only 35.71\% precision and 41.67\% recall, with its F1 score decreasing to 38.46\%. Gemini 3.1 Pro retains relatively high conditional recall at 81.82\% and improves its F1 score to 58.06\%, but only 41 of its 50 responses are valid. Its full-set agreement therefore decreases slightly to 56.00\%. Opus 5.5 remains comparatively balanced, achieving 64.29\% precision and 69.23\% recall, although its F1 score and full-set agreement decrease to 66.67\% and 80.00\%, respectively.
\section{Conclusion}
In this paper, we present StarCore, a cybersecurity world model that advances large language models toward dynamic understanding and reasoning of cyberspace. StarCore introduces a unified system state-chain representation to capture the temporal and behavioral relationships among heterogeneous data, including code execution, network traffic, system commands, and agent interactions. Building on this representation, we develop a two-stage learning framework that combines state-chain supervised fine-tuning with GRPO-based reinforcement learning, enabling the model to understand system states, reason about behavioral dependencies, predict potential attacks, and perform security assessments. Extensive experiments demonstrate that StarCore achieves strong performance across diverse cybersecurity tasks. In particular, StarCore-180B achieves an overall answer accuracy of 0.9238, outperforming the strongest baseline by 16.77 percentage points, while the compact StarCore-9B achieves the highest reasoning correctness score of 52.8 and surpasses much larger models in security assessment. These results highlight the effectiveness of state-chain modeling and learning for cybersecurity world models.


\bibliographystyle{abbrv}
\bibliography{ref}

\clearpage
\appendix

\section*{Contribution}
\label{sec:contribution}

The listing of contributors is in alphabetical order based on their last names.

\noindent \textbf{Core Contributors}

\begin{multicols}{4}
\noindent
Tianyu Cui\\  Miao Chen\\  Yang Li\\ Ziwei Li\\ Jiaxin Shi\\  Ao Wang\\ Shouyi Wang\\ Weijie Wu\\  Yue Xiao\\  Min Zhu\\ Yixiang Zhang
\end{multicols}

\textbf{Technique Advisors}

\noindent Qi Li \hspace{2.65cm} Ke Xu

\clearpage
\section{Notation and Formal Problem Statement}
\label{app:formalization}

This appendix consolidates the notation used throughout the paper and explicitly distinguishes latent system dynamics, observed artifacts, and model predictions. 

\subsection{Notation}
\label{app:notation}

\begin{table}[h]
\centering
\caption{Summary of notation. }
\label{tab:app-notation}
\small
\setlength{\tabcolsep}{5pt}
\resizebox{\textwidth}{!}{
\begin{tabularx}{\linewidth}{@{}lX@{}}
\toprule
\textbf{Symbol} & \textbf{Meaning} \\
\midrule
$t, T$ & Discrete time index and final index of an observed sequence. \\
$o_t$ & Observed system artifact at time $t$, such as a code fragment, flow record, or command event. \\
$a_t$ & Latent or annotated system action associated with a transition. \\
$s_t$ & Canonical state representation used by the state-chain model. \\
$S=\{s_t\}_{t=0}^{T}$ & A uni-modal or synchronized system state chain. \\
$\widetilde{S}$ & A partially observed chain in which selected states are masked. \\
$\mathcal{M}$ & Set of indices of missing states in a completion sample. \\
$C$ & Environmental context, including the system or session context available to the model. \\
$Q,\mathcal{I}$ & Task-specific question and system instruction. \\
$\tau$ & Task type: completion, prediction, or security assessment. \\
$R$ & Action reasoning or an evidence-linked intermediate rationale. \\
$Y$ & Final structured task output. \\
$K$ & Prediction horizon measured in future states. \\
$x$ & Complete model input $(\mathcal{I},Q,S,C,\tau)$. \\
$o=(R,Y)$ & A generated response trajectory; its meaning is distinct from an observation $o_t$. \\
$\pi_\theta,\pi_{\mathrm{ref}}$ & Trainable policy and frozen reference policy. \\
$G$ & Number of responses sampled for one GRPO rollout group. \\
$R_{\mathrm{SCR}}$ & Aggregated State Chain Reward. \\
\bottomrule
\end{tabularx}}
\end{table}

\subsection{Observations, States, and Transitions}
\label{app:observation-model}

Let the latent evolution of a system be written as
\begin{equation}
    s^{\star}_{t+1}\sim P_{\mathrm{env}}\left(s^{\star}_{t+1}\mid s^{\star}_{t},a_t\right),
    \qquad
    o_t\sim P_{\mathrm{obs}}\left(o_t\mid s^{\star}_t\right),
    \label{eq:app-latent-transition}
\end{equation}
where $s^{\star}_t$ denotes the generally unobserved true system state. StarCore does not assume that a single artifact fully determines $s^{\star}_t$. Instead, raw observations are normalized into model-facing states $s_t=\phi(o_{\le t}, C)$, where $\phi$ denotes deterministic parsing, canonicalization, and context attachment. A state chain is therefore an evidence sequence rather than a claim of complete observability.

For training, the action $a_t$ may be unavailable. In this case, the learning target is the observed transition regularity $P(s_{t+1}\mid s_{\le t}, C)$, and action reasoning is treated as an explanation constrained by available evidence. When reliable action annotations exist, they can provide additional supervision, but they should remain distinguishable from model-generated rationales in the released data format.

\subsection{Uni-Modal Chains and Interaction Graphs}
\label{app:graph-definition}

A uni-modal chain of modality $m$ is an ordered sequence
\begin{equation}
    c^{(m)}=\left(s^{(m)}_0,s^{(m)}_1,\ldots,s^{(m)}_T\right),
    \qquad
    (s^{(m)}_t,s^{(m)}_{t+1})\in E_{\mathrm{temp}},
\end{equation}
where $E_{\mathrm{temp}}$ contains edges justified by execution order or event time. A multi-modal interaction graph is $\mathcal{G}=(\mathcal{V},\mathcal{E})$, with $\mathcal{V}$ equal to the union of canonical states and
\begin{equation}
    \mathcal{E}=E_{\mathrm{temp}}\cup E_{\mathrm{align}}\cup E_{\mathrm{dep}}.
\end{equation}
Here, $E_{\mathrm{align}}$ links observations of the same underlying event across modalities, while $E_{\mathrm{dep}}$ records explicitly supported behavioral dependencies. These edge types should not be conflated, i.e., temporal adjacency alone does not establish causality, and a cross-modal alignment edge does not by itself prove malicious intent.

\section{State-Chain Construction and Data Quality Control}
\label{app:construction}

This section gives a reference protocol for converting heterogeneous artifacts into state chains. It separates deterministic data transformations from semantic annotations and identifies the checks needed to make the resulting samples auditable. Implementation-specific thresholds, when used, should be recorded in dataset cards rather than hidden in preprocessing code.

\subsection{Construction Pipeline}
\label{app:construction-pipeline}

\begin{algorithm}[h]
\caption{Reference state-chain construction procedure}
\label{alg:app-chain-construction}
\begin{algorithmic}[1]
\REQUIRE Raw artifact collection $\mathcal{D}_{\mathrm{raw}}$ and source metadata
\ENSURE Validated state-chain samples with provenance
\STATE Preserve source identifiers, collection time, licenses, and integrity hashes.
\STATE Parse each artifact into atomic events; retain unparsable records in a rejection log.
\STATE Canonicalize timestamps, identifiers, field types, and redacted values without discarding the source mapping.
\STATE Partition events into sessions using modality-specific chain keys.
\STATE Sort events by execution order when available, otherwise by timestamp with an explicit uncertainty flag.
\STATE Form uni-modal chains and attach the environmental context available at each cutoff.
\STATE Add cross-modal links only when shared identifiers, aligned sessions, or other documented evidence support them.
\STATE Validate ordering, schema conformity, target isolation, and train--test separation.
\STATE Return accepted chains, rejection reasons, and aggregate quality reports.
\end{algorithmic}
\end{algorithm}

The rejection log is part of the data pipeline rather than disposable debugging output. It enables users to distinguish absence of malicious behavior from absence of usable observations. Similarly, redaction mappings should be stable within a chain so that entity consistency is preserved even when raw identifiers cannot be released.

\subsection{Modality-Specific State Units}
\label{app:modality-units}

Canonicalization should remove serialization differences that are irrelevant to the task while preserving security-relevant distinctions. For example, equivalent whitespace can be normalized in structured records, but command arguments, exit status, traffic direction, and tool-call errors must not be discarded. When multiple raw events map to one canonical state, the mapping should be many-to-one and traceable rather than silently overwriting observations.

\subsection{Temporal Alignment and Cross-Modal Linking}
\label{app:alignment}

Clock skew and collection latency make exact timestamp equality an unreliable linking rule. A defensible alignment record should contain the linked state identifiers, the evidence used for linking, and an uncertainty indicator. Shared process or session identifiers provide stronger evidence than timestamp proximity alone. If only time windows are available, the link should be treated as uncertain and excluded from labels that require definitive cross-modal causality.

A cutoff-safe representation is essential for prediction. For a sample with cutoff $t$, the input construction function must satisfy
\begin{equation}
    \mathrm{Input}(t)=\mathrm{Filter}\left(\mathcal{D}_{\mathrm{raw}},\; \mathrm{available\_at}\le t\right),
\end{equation}
and must exclude future flow summaries, post-incident annotations, final command outcomes, and agent reflections produced after $t$. This restriction prevents subtle label leakage even when the visible future states themselves are masked.

\subsection{Missingness and Chain Integrity}
\label{app:missingness}

Completion targets should be generated from intact source chains before masking. Let $M\in\{0,1\}^{T+1}$ denote a mask with $M_t=1$ for a hidden state. The corrupted chain is
\begin{equation}
    \widetilde{s}_t=(1-M_t)s_t+M_t\,\texttt{[MASK]},
    \qquad t\in\{0,\ldots,T\}.
\end{equation}
Masks should be stored separately from the prompt so that evaluation can distinguish random, contiguous, and modality-specific missingness. Naturally missing observations should not automatically be treated as completion targets unless their values can be recovered from an independent source.

Chain validation covers four properties: (i) \emph{well-formedness}, meaning that every state conforms to the schema; (ii) \emph{order validity}, meaning that sequence indices and event times do not conflict; (iii) \emph{referential validity}, meaning that cross-modal links resolve to existing states; and (iv) \emph{target isolation}, meaning that no answer field is copied into an input field. Samples that fail these checks should be repaired deterministically or excluded with a logged reason.

\section{State Chain Reward and GRPO Implementation Details}
\label{app:rewards}

This appendix expands the reward definitions in Section~\ref{sec:scr} into verifiable inputs, task-dependent activation rules, and optimization diagnostics. It does not introduce additional reward weights or training hyperparameters; those values should be reported with the corresponding training configuration.

\subsection{Verifier Inputs and Failure Isolation}
\label{app:verifiers}

Each reward component should be computed from declared evidence rather than from an unconstrained judgment of the whole response. Table~\ref{tab:app-verifiers} lists a reference verifier contract. Programmatic checks are preferable for schema validity, state indices, and exact state reconstruction. Learned or human judges, when used for semantic components, should be calibrated separately and their agreement reported.

\begin{table*}[t]
\centering
\caption{Reference verifier contract for State Chain Reward components.}
\label{tab:app-verifiers}
\small
\setlength{\tabcolsep}{5pt}
\begin{tabularx}{\textwidth}{@{}lXX X@{}}
\toprule
\textbf{Component} & \textbf{Required evidence} & \textbf{Typical checks} & \textbf{Failure isolated by the component} \\
\midrule
Format & Response and task schema & Parseability, required fields, valid indices, label vocabulary & Structurally invalid or incomplete output. \\
State understanding & Visible states and parsed response entities & Entity consistency, direction, event type, temporal references & Misreading an otherwise well-formatted observation. \\
Causal reasoning & Visible states, rationale, optional action annotations & Evidence coverage, transition consistency, unsupported inference & Correct label supported by an invalid explanation. \\
Attack prediction & Pre-cutoff input, predicted states, reference future & Horizon validity, future-state agreement, calibrated alternatives & Plausible narrative inconsistent with the actual future. \\
Security assessment & State evidence, rationale, final assessment & Label agreement, evidence sufficiency, benign alternatives & Unsupported attribution or missed malicious behavior. \\
\bottomrule
\end{tabularx}
\end{table*}

Let $v_k(x,o)$ denote the raw output of verifier $k$. A bounded reference normalization is
\begin{equation}
    r_k(x,o)=\frac{\min(\max(v_k(x,o),l_k),u_k)-l_k}{u_k-l_k},
    \qquad u_k>l_k,
    \label{eq:app-reward-normalization}
\end{equation}
where $[l_k,u_k]$ is the declared raw score range. If a verifier is already bounded in $[0,1]$, no additional normalization is required. Scores from different verifier types should not be combined until their ranges and missing-value conventions have been made explicit.

\subsection{Task-Dependent Reward Activation}
\label{app:reward-masks}

\begin{table}[h]
\centering
\caption{Reference activation mask $m_{\tau,k}$. An entry of 1 denotes an applicable component; 0 denotes an inactive component. Active weights are renormalized for each task as in the main text.}
\label{tab:app-reward-mask}
\small
\setlength{\tabcolsep}{6pt}
\begin{tabular}{@{}lccccc@{}}
\toprule
\textbf{Task} & \textbf{Format} & \textbf{State} & \textbf{Causal} & \textbf{Prediction} & \textbf{Assessment} \\
\midrule
Completion & 1 & 1 & 1 & 0 & 0 \\
Prediction & 1 & 1 & 1 & 1 & 0 \\
Security assessment & 1 & 1 & 1 & 0 & 1 \\
\bottomrule
\end{tabular}
\end{table}

For a task $\tau$, the active set is $\mathcal{K}_{\tau}=\{k:m_{\tau,k}=1\}$ and
\begin{equation}
    \widetilde{w}_{\tau,k}=\frac{m_{\tau,k}w_k}{\sum_{j\in\mathcal{K}}m_{\tau,j}w_j},
    \qquad
    R_{\mathrm{SCR}}^{(\tau)}(x,o)=\sum_{k\in\mathcal{K}}\widetilde{w}_{\tau,k}r_k(x,o).
    \label{eq:app-active-reward}
\end{equation}
If no component is active, the sample is configuration-invalid and should not enter training. A malformed response receives a low format score, but we must decide in advance how to treat its remaining components. Assigning zero to all semantic components is stricter. Therefore, evaluating whatever can be parsed is more diagnostic. Mixing these conventions across runs would make reward comparisons difficult to interpret.

\subsection{GRPO Update Procedure}
\label{app:grpo-procedure}

\begin{algorithm}[h]
\caption{One state-chain GRPO optimization step}
\label{alg:app-grpo-step}
\begin{algorithmic}[1]
\REQUIRE Policy $\pi_\theta$, reference policy $\pi_{\mathrm{ref}}$, batch $\mathcal{B}$, group size $G$
\ENSURE Updated policy parameters and reward diagnostics
\FOR{each input $x\in\mathcal{B}$}
\STATE Sample $\{o_i\}_{i=1}^{G}$ from the rollout policy $\pi_{\theta_{\mathrm{old}}}$.
\STATE Evaluate active SCR components and store component-level scores.
\STATE Compute $r_i=R_{\mathrm{SCR}}(x,o_i)$ for every candidate.
\STATE Compute $\mu_r$ and $\sigma_r$ within the rollout group.
\IF{$\sigma_r$ is below the declared numerical threshold}
\STATE Set group advantages to zero and mark the group as degenerate.
\ELSE
\STATE Compute $\hat{A}_i=(r_i-\mu_r)/(\sigma_r+\epsilon)$.
\ENDIF
\ENDFOR
\STATE Maximize the clipped objective with KL regularization over valid groups.
\STATE Log reward distributions, format failures, KL statistics, and degenerate-group rates.
\end{algorithmic}
\end{algorithm}

A degenerate group contains candidates with identical or nearly identical rewards. Its relative advantages carry no useful ordering signal. We set them to zero to avoid amplifying numerical noise. The cause should still be inspected. Repeated degenerate groups may indicate that prompts are too easy, sampling diversity is too low, or the verifier cannot distinguish meaningfully different responses.

\subsection{Reward Diagnostics and Anti-Gaming Checks}
\label{app:reward-diagnostics}

A scalar training reward can rise while one underlying capability deteriorates. Component-level curves should therefore be reported alongside the aggregate SCR. Useful diagnostics include format validity, average evidence coverage, future-state accuracy, assessment accuracy, response length, KL divergence, and the fraction of degenerate groups. Abrupt changes in only one component should trigger a verifier audit before being interpreted as improved world understanding.

The verifier itself should be tested with counterexamples: a correct answer paired with fabricated evidence; a well-formatted response that cites invisible states; a prediction copied from post-cutoff context; a benign chain labeled malicious without support; and a verbose rationale that repeats observations without explaining a transition. These probes help detect reward hacking and distinguish structural validity from semantic correctness. Human review should focus on samples where the answer and rationale rewards disagree, since those cases expose failures hidden by final-answer accuracy alone.

\end{document}